\documentclass[32pt, usenatbib]{mn2e}
 \usepackage{graphicx,times,subfigure}
\usepackage{bm}
\usepackage{epsf}
\usepackage{amsmath}
\usepackage{amssymb}
\usepackage{cases}
\usepackage{float}
\usepackage{colortbl}
\usepackage{color}
\usepackage{multirow}
\usepackage{url}
\usepackage{hyperref}

\def\cb{\textcolor{black}}

\def\la{\langle}
\def\ra{\rangle}
\def\n{\noindent}
\def\be{\begin{equation}}
\def\ee{\end{equation}}
\def\ben{\begin{eqnarray}}
\def\een{\end{eqnarray}}
\def\nn{\nonumber}
\def\oh{\bf \hat\Omega}

\def\bk{{\bf k}}

\def\ad{{_{\rm{lin}}}}
\def\bk{{\bf k}}

\def\bl{{\bf l}}

\def\2p{{(2\pi)^2}}

\def\bl{{\bf l}}
\def\be{\begin{equation}}
\def\ee{\end{equation}}
\def\beq{\begin{equation}}
\def\eeq{\end{equation}}
\def\ben{\begin{eqnarray}}
\def\een{\end{eqnarray}}
\def\bes{\begin{subequations}}
\def\ees{\end{subequations}}

\def\oh{{\hat\Omega}}
\def\nn{{\nonumber}}

\newcommand{\beqa}{\begin{eqnarray}}
\newcommand{\eeqa}{\end{eqnarray}}

\def\ikap0{{\cal J}_{\theta_0}(r)}

\def\one1{\langle \kappa_{(i)}\kappa_{(j)} \rangle}
\def\one{{[\bar \xi^{(ij)}]}}

\def\ba{\begin{eqnarray}}
\def\ea{\end{eqnarray}}

\def\bk{{\bf k}}

\def\n{\noindent}

\def\ad{{_\delta}}

\def\bk{{\bf k}}

\def\bl{{\bf l}}

\def\2p{{(2\pi)^2}}

\def\bl{{\bf l}}

\def\be{\begin{equation}}
\def\ee{\end{equation}}

\def\beq{\begin{equation}}
\def\eeq{\end{equation}}

\def\ben{\begin{eqnarray}}
\def\een{\end{eqnarray}}

\def\oh{{\hat\Omega}}
\def\nn{{\nonumber}}

\def\n{\noindent}

\def\ad{{_\delta}}

\def\bk{{\bf k}}

\def\bl{{\bf l}}

\def\2p{{(2\pi)^2}}

\def\bl{{\bf l}}

\def\bl{{\bf{l}}}

\def\btheta{{\boldsymbol\theta}}

\title[Weak Lensing Skew-Spectrum]
      {Weak Lensing Skew-Spectrum}
      \author[Munshi et al.]
             {D. Munshi$^a$, T. Namikawa$^b$, T. D. Kitching$^a$, J. D. McEwen$^a$,
               F. R. Bouchet$^c$\\
               $^{a}$ Mullard Space Science Laboratory, University College London,
               Holmbury St Mary, Dorking, Surrey RH5 6NT, UK \\
               $^{b}$ Department of Applied Mathematics and Theoretical Physics,
               University of Cambridge, Wilberforce Road, Cambridge CB3 OWA, UK \\
               $^{c}$ Institut d’Astrophysique de Paris, UMR 7095, CNRS \& Sorbonne Universit,
               98 bis Boulevard Arago, F-75014 Paris, France \\
               }
\begin{document}
\onecolumn
\maketitle
%
%
\begin{abstract}
{We introduce the skew-spectrum statistic for weak lensing 
  convergence $\kappa$ maps and test it against state-of-the-art high-resolution all-sky numerical simulations.
  We perform the analysis as a function of source redshift and smoothing angular scale for individual
  tomographic bins. We also analyse the cross-correlation between different tomographic bins.
  We compare the numerical results to fitting-functions 
  used to model the bispectrum of
  the underlying density field as a function of redshift and
  scale. We derive a closed form expression for the skew-spectrum
  for gravity-induced secondary non-Gaussianity.
  We also compute the skew-spectrum for the
  projected $\kappa$ inferred from Cosmic Microwave Background (CMB) studies.
  As opposed to the low redshift case we find the post-Born corrections to be important
  in the modelling of the skew-spectrum for such studies.
  We show how the presence of a mask and noise can be incorporated in the estimation of a skew-spectrum.}
\end{abstract}
%
%
\section{\bf Introduction}
\label{sec:intro}
%
%
Recently completed CMB experiments
such as the
Planck Surveyor\footnote{{\href{http://http://sci.esa.int/planck/}{\tt http://http://sci.esa.int/planck/}}}\citep{Planck1,Planck18},
have helped establishing a standard model of cosmology, with the baseline cosmological parameters
now known with an unprecedented accuracy.
However, many fundamental questions in cosmology remain open. These include the nature of dark matter
and dark energy, a possible modification of
General Relativity on cosmological scales \citep{MG1,MG2}
and the nature of neutrinos mass hierarchy\citep{nu}.
Next generation of large scale surveys will provide a massive amount of high-precision data
carrying complementary information that can help answer at least some of these questions.
Indeed, observational programs of many ongoing as well as future surveys
including the surveys e.g.,
{\em Euclid}\footnote{\href{http://sci.esa.int/euclid/}{\tt http://sci.esa.int/euclid/}}\citep{Euclid}, 
CFHTLS\footnote{\href{http://www.cfht.hawai.edu/Sciences/CFHLS/}{\tt http://www.cfht.hawai.edu/Sciences/CFHLS}},
PAN-STARRS\footnote{\href{http://pan-starrs.ifa.hawai.edu/}{\tt http://pan-starrs.ifa.hawai.edu/}},
Dark Energy Surveys\footnote{\href{https://www.darkenergysurvey.org/}{\tt https://www.darkenergysurvey.org/}}\citep{DES},
WiggleZ\footnote{\href{http://wigglez.swin.edu.au/}{\tt http://wigglez.swin.edu.au/}}\citep{WiggleZ},
Rubin Observatory,\footnote{\href{http://www.lsst.org/llst home.shtml}{\tt {http://www.lsst.org/llst home.shtml}}}\citep{LSST_Tyson},
BOSS\footnote{\href{http://www.sdss3.org/surveys/boss.php}{\tt http://www.sdss3.org/surveys/boss.php}}\citep{SDSSIII},
KIDS\citep{KIDS}, Roman Space Telescope\citep{WFIRST}, lists weak lensing as their main science driver.
From the early days of detection weak lensing (see e.g. \citep{review} for a review) studies have now reached
a level of maturity. Surveys such as {\em Euclid} will constrain the cosmological
parameters with sub-percent accuracy and answer many of the most challenging questions that cosmology
is facing today.
 
Weak lensing at smaller angular scales probes scales that are in the highly nonlinear regime
and contains a wealth of cosmological information. This gravity-induced nonlinearity \citep{bernardeau_review}
introduces mode-coupling that is responsible for the resulting departure from Gaussianity \citep{review_ng}.
Higher-order statistics beyond power spectrum estimation is typically used in exploitation of the
information content of weak lensing maps. An accurate modelling of higher-order
statistics is important for modelling the
covariance of the lower order estimators as well as to break cosmological parameter degeneracy.
Early studies of higher-order statistics concentrated on cumulants \citep{Bernard1,Bernard2}
in real-space \citep{Mellier1,Mellier2}. Future surveys such as {\em Euclid} will have a near all-sky coverage
and thus enable quantifying higher-order statistics in the harmonic domain where
measurements of individual modes will be less correlated \citep{Euclid_Review}. 

Most theoretical modelling in the highly nonlinear regime were based on perturbative
calculations or its extensions \citep{bernardeau_review}, variants of the halo models \citep{CooraySheth},
Effective Field Theory \citep{EFT1} or fitting-functions that are calibrated from simulations \citep{Scoccimarro,GM11}.
Many different estimators are currently available for analysing departures from Gaussianity,
including morphological estimators \citep{Waerbeke}, position-dependent power
spectra \citep{IB}, line-correlations \citep{ES17}, 
extreme value statistics \citep{HC11}.
peak-statistics \citep{peak1,peak2}, void statistics \citep{void1} and
probability distribution functions \citep{PDF1,PDF2,PDF3,PDF4}.

While measurements of real space correlations are much simpler in
the presence of complicated survey boundaries the measurements for different 
angular scales can be highly correlated \citep{MunshiJain1,MunshiJain2,Munshi_bias}. In comparison the measurements
in the harmonic domain are less correlated and contains independent information if the sky coverage is high.
One of the motivation of this study is to develop 
analytical predictions for one such proxy statistics to the bispectrum called skew-spectrum
\citep{MunshiHeavens,Cooray01} and test them against state-of-art numerical simulations.
We will borrow the concepts developed for constructing skew-spectrum for the the study of
non-Gaussianity in the context of Cosmic Microwave Background (CMB) observations by
WMAP\footnote{\href{https://map.gsfc.nasa.gov/}{\tt https://map.gsfc.nasa.gov/}}\citep{Smidt1,Erminia}
and {\em Planck}\citep{Planck_NG} satellites.
However our aim here is also to include gravity induced secondary non-Gaussianity.
The skew-spectrum is the lowest-order member in the family of higher-order spectra \citep{xn1,xn2}.
They can also be used to reconstruct morphological estimators,
e.g., Minkowski Functionals, in an order by order manner in
the presence of complicated survey topology \citep{Waerbeke}.
Recently the skew-spectrum statistics was used to study the possibility of probing galaxy clustering using
data from the forthcoming generation of wide-field galaxy surveys\citep{Dai02,Seljak15,Cora20}.

\cb{In this paper we show that the skew-spectrum statistics can be used to analyse the
weak lensing maps that will be available from the future stage-IV experiments such as
{\em Euclid} or the Rubin Observatory. We also show how the sub-optimal skew-spectrum
can be used to probe the gravity-induced non-Gaussianity of the reconstructed
convergence maps from CMB observations. We will present the skew-spectrum
by cross-correlating different tomographic bins as well as  the CMB convergence maps and the low redshift weak lensing convergence maps.
In this context we will emphasize the importance of the post-Born corrections
in theoretical modeling of the bispectrum. Finally, we will consider many modified theories
of gravity and use second-order perturbation theory to model
the theoretical skew-spectrum at large smoothing angular scales to provide
an example of important science goals that can be achieved using the skew-spectrum
statistics.}

This paper is organised as follows.
In \textsection\ref{sec:bispec} we briefly review the modelling of the
density bispectrum. In \textsection\ref{sec:flat_sky}
we introduce our notations and briefly summarize the results
of projected weak lensing convergence or $\kappa$ bispectrum.
The \textsection\ref{sec:simu} is devoted to the discussion of
the simulations we use in our study. 
In \textsection\ref{estimator} we present the estimator we use.
The results are discussed in \textsection\ref{sec:disc} and
conclusion and future prospects are presented in \textsection\ref{sec:conclusion}.

\section{Modelling of the Density Bispectrum}\label{sec:bispec}
%
In this section we relevant the aspects of tree-level perturbative
results and their extensions using approaches based on fitting-function,
which we use to compute the bispectrum and eventually the skew-spectrum.

\subsection{Tree-level Perturbative Calculations}
\label{subsec:tree}
%
\begin{figure}
  \begin{center}
  \begin{minipage}[b]{0.45\textwidth}
    \includegraphics[width=\textwidth]{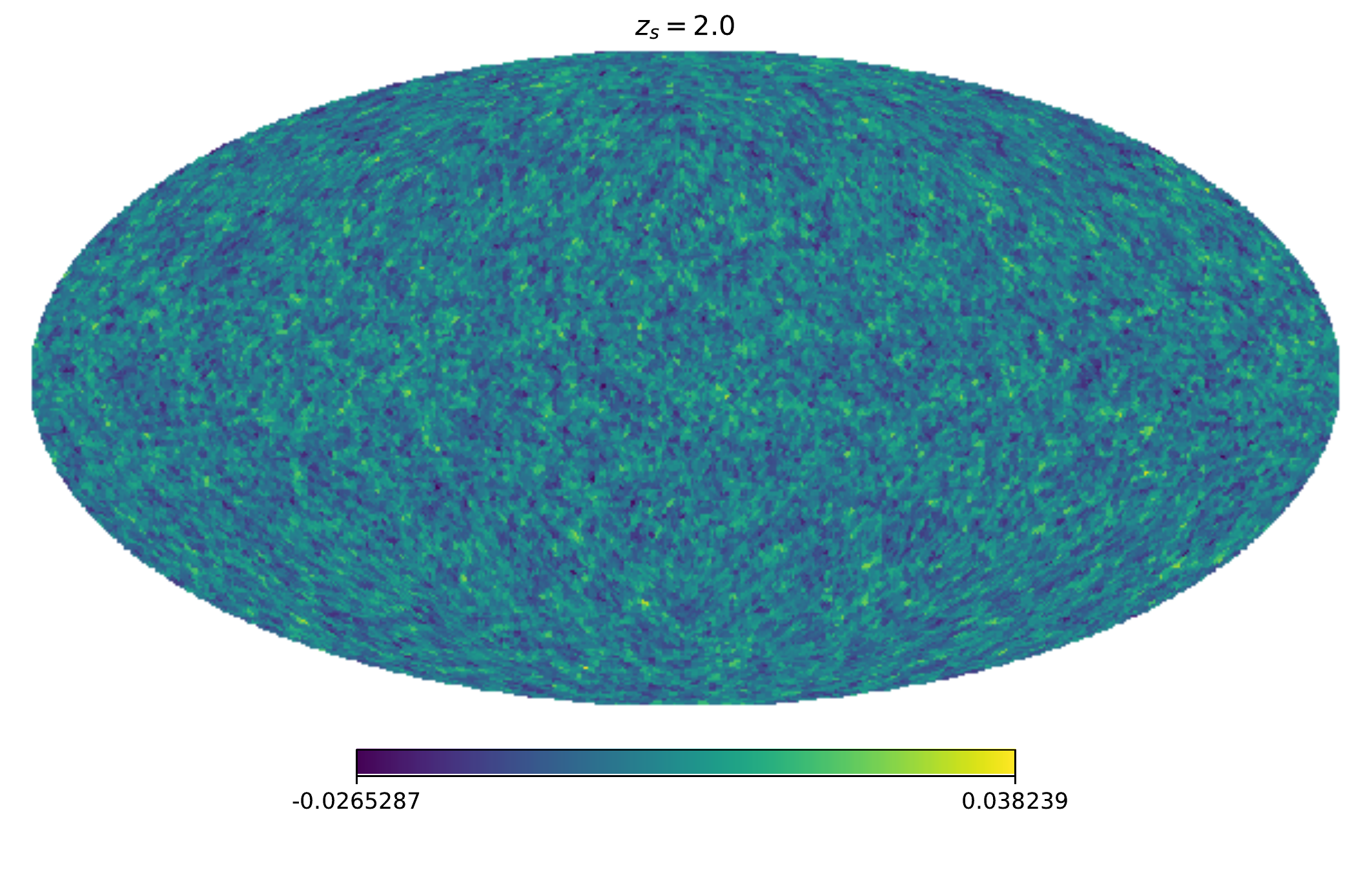}
    \label{fig:1}
  \end{minipage}
  \begin{minipage}[b]{0.45\textwidth}
    \includegraphics[width=\textwidth]{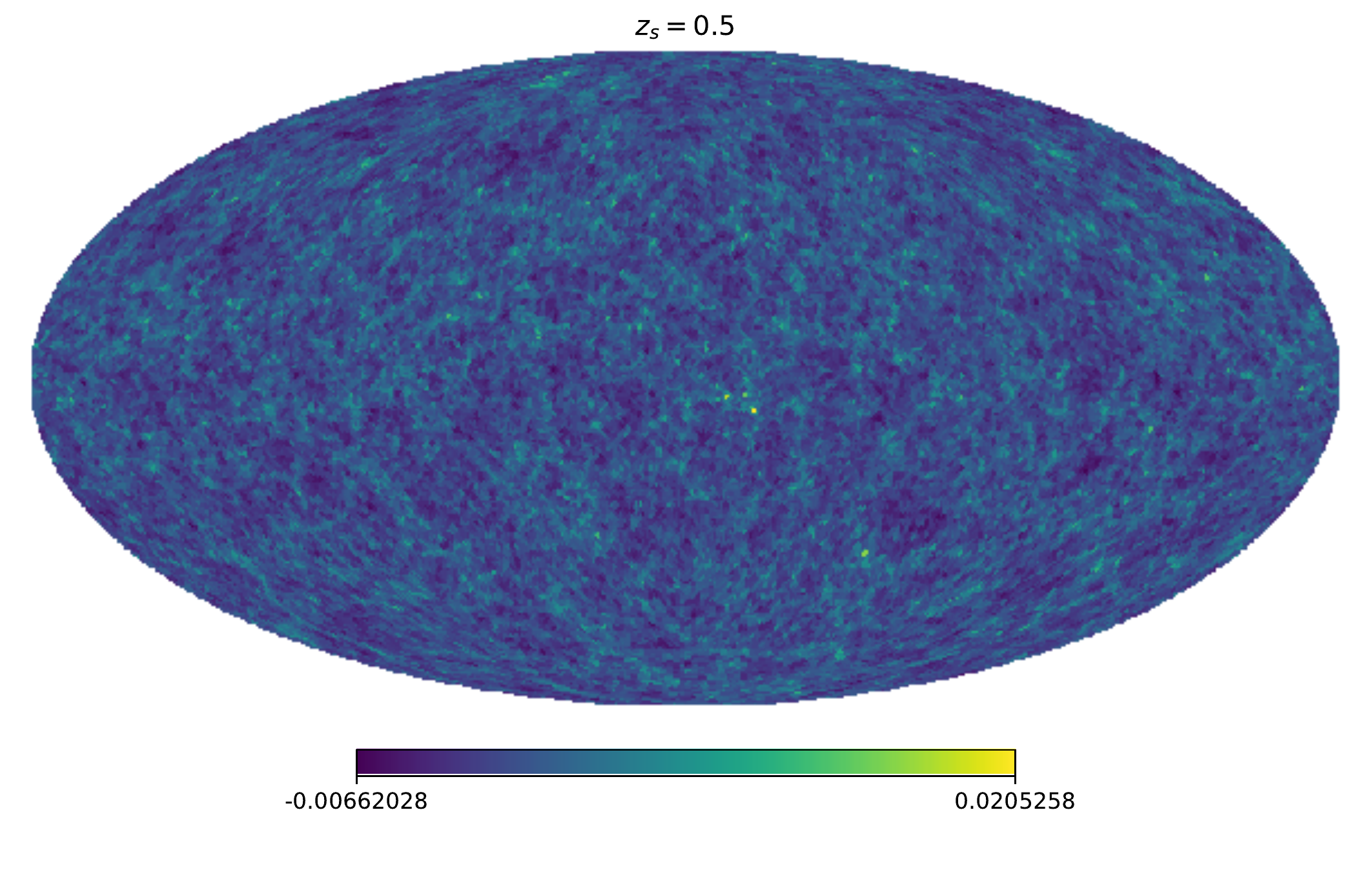}
    \label{fig:2}
  \end{minipage}
  \caption{Examples of realisations of all-sky weak lensing convergence or $\kappa$ maps used for our study.
    The left- and right-panel panels correspond respectively 
    source redshift $z_s=2.0$ and $0.5$. The $\kappa$ maps, we have used, were generated at
    a Healpix resolution $N_{\rm side}=4096$, and we have degraded them
    to $N_{\rm side}=2048$ for our study.}
  \label{fig:allsky}
  \end{center}
\end{figure}
%
In the weakly non-linear regime with density contrast ($\delta \le 1$), the
gravitational clustering can be described by the Eulerian perturbation theory
\cite{review}. However, the perturbative treatment eventually breaks
down when density contrast at a given length scale becomes nonlinear
($\delta \ge 1$). Expanding the density contrast $\delta$ in a Fourier
series, and assuming the density contrast is less than unity, for
the pertubative series to be convergent, we get
\bes\ben
&&\delta({\bf k}) = \delta^{(1)}({\bf k}) + \delta^{(2)}({\bf k})
+ \delta^{(3)}({\bf k}) + \dots; \\
&&\delta^{(2)}(k) = \int {
d^3\bk_1 \over 2\pi} \int { d^3 \bk_2 \over 2\pi} \delta_D({\bf k_1 +
k_2 -k }) F_2(\bk_1,\bk_2) \delta^{(1)}({\bf k}_1) \delta^{(1)}({\bf
  k}_2); \label{eq:F2}\\
&& F_2(\bk_1,\bk_2)={5\over 7} +{1\over 2}\left({k_1\over k_2}+{k_2\over k_1}\right ) 
\left({\bk_1\cdot\bk_2 \over k_1 k_2}\right )
+{2 \over 7}
\left( {\bk_1\cdot\bk_2 \over k_1 k_2}\right )^2.
\label{eq:F2}
\een\ees
\n The linearized solution for the density field is
$\delta^{(1)}({\bf k})$; higher-order terms $\delta^{(2)}, \delta^{(3)}, \cdots$ yield second- and third-order corrections to
this linear solution. The 3D wave vectors are denoted as ${\bk}, {\bk}_1,{\bk}_2$ and their magnitudes
as $k=|\bk|$ and $k_i = |{\bk}_i|$.
More details of our Fourier convention will be introduced in \textsection\ref{sec:new_allflat}.
Using a fluid approach known to be valid at
large scales (and before shell crossing) one can write the second
order correction to the linearized density field using the kernel
$F_2({\bf k_1},{\bf k_2})$. While we will be taking the fitting-function approach,
in recent years many new methods have been developed to tackle the gravitational instability in the nonlinear regime
including Effective Field Theory (EFT) methods (see e.g. \cite{MunshiRegan} and references therein).
%
%
\subsection{Phenomenological Fitting-Functions}\label{sec:eftoflss}
%
Beyond the quasilinear regime 
non-perturbative tools become necessary. One such approach was developed in
\citep{Scoccimarro} who proposed the so-called Hyper Extended Perturbation
Theory (HEPT) in the highly nonlinear regime and a fitting-function that connects it with the tree-level perturbative
calculation. The fitting function which interpolates these two regime is calibrated using numerical simulations.
Over the years similar but  more accurate fitting formula were developed by other authors \citep{GM11},
which essentially generalise the kernel $F_2$ defined in Eq.(\ref{eq:F2})
by introducing  scale-dependent coefficients $a(n_i,k_i),b(n_j,k_j)$ and $c(n_j,k_j)$:
\ben
&& F_2(\bk_i,\bk_j)={5\over 7}a(n_i,k_i)a(n_j,k_j) +
    {1\over 2}\left( {\bk_i\cdot\bk_j \over k_i k_j}\right )\left({k_i\over k_j}
    +{k_j\over k_i}\right ) b(n_i,k_i) b(n_j,k_j) \nn \\
    &&  \hspace{2cm}  + {2 \over 7}\left( {\bk_i\cdot\bk_j \over k_i k_j}\right )^2 \,c(n_i,k_i)\, c(n_j,k_j).
    \label{eq:fit1}
    \een
Here $n_{e}$ is the effective spectral slope associated with the
linear power spectra $n_{e} = d \ln P_\ad(k) / d\ln k $, $q$ is the ratio
of a given length scale to the non-linear length scale $q=k/k_{\rm nl}$,
where ${k_{\rm nl}^3/2\pi^2}D_+^2(z)P\ad(k_{\rm nl}) = 1$.
Here $D_+(z)$ represents the linear growth rate of perturbations at redshift $z$. 
At length scales where $q \ll 1$, 
the relevant length scales are well within the quasilinear
regime, $a=b=c=1$, and we recover the tree-level perturbative results.
In the regime where $q \gg 1$, and the length scales we are
considering are well within the nonlinear scale, we recover $a \ne 1 $ but $b=c=0$. In this
limit the bispectrum becomes independent of configuration.
It was recently pointed
out by \citep{MunshiNamikawa19} that the fitting function of \citep{GM11}
is not very accurate in describing the weak lensing bispectrum.
A more accurate fitting function was developed  recently in \citep{Taka19}
which we will be using in this study. The original fitting function by
\citep{Scoccimarro} involved just six free parameters and was valid
for $k< 3h {\rm Mpc}^{-1}$ and $0< z< 1$. The improved \citep{GM11} formula
has a limited range of validity $k <0.4 h {\rm Mpc}^{-1}$ and $0<z<1.5$
and contains nine parameters. The fitting function by \citep{Taka19}
contains $52$ free parameters. Introduction of such a large number of
free parameters increases the validity range to $k=10 h{\rm Mpc}^{-1}$ and $z \approx 1-3$.
This will be important in modeling non-Gaussianity on arcminute scales probed by
the future stage-IV experiments.
%
\begin{figure}
  \begin{center}
  \begin{minipage}[b]{0.45\textwidth}
    \includegraphics[width=\textwidth]{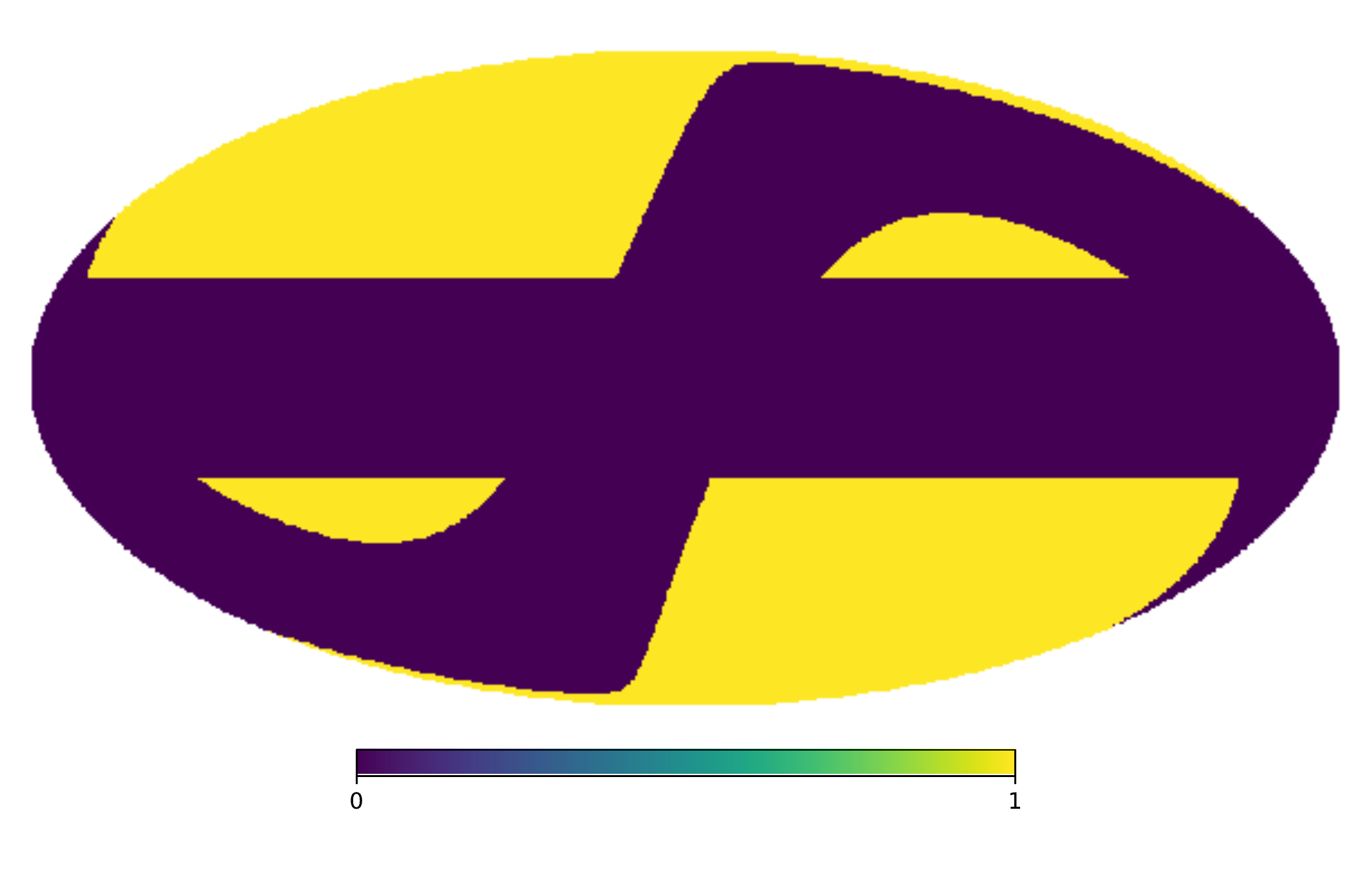}
  \end{minipage}
  \caption{In our study, we use a ``pseudo Euclid'' mask. In constructing the mask,  all pixels (shown in dark) lying within $22$ degree of either
    the galactic or ecliptic planes are discarded. The remaining unmasked pixels (shown in yellow) cover $14,490$ degree$^2$
    of the sky, making fraction of the sky covered $f_{\rm sky} \approx 0.35$.}
   \label{fig:myhealpymask}
  \end{center}
\end{figure}
%
%
\section{Weak Lensing Statistics in Projection}
\label{sec:flat_sky}
In this section we will relate the convergence bispectrum $\kappa$
with its 3D density contrast $\delta$ counterpart. Then this bispectrum
will be used to construct the convergence skew-spectrum.
\subsection{Projected Weak lensing Bispectrum}\label{sec:skewspec}
%
\begin{figure}
  \centering
  \includegraphics[width=0.9\textwidth]{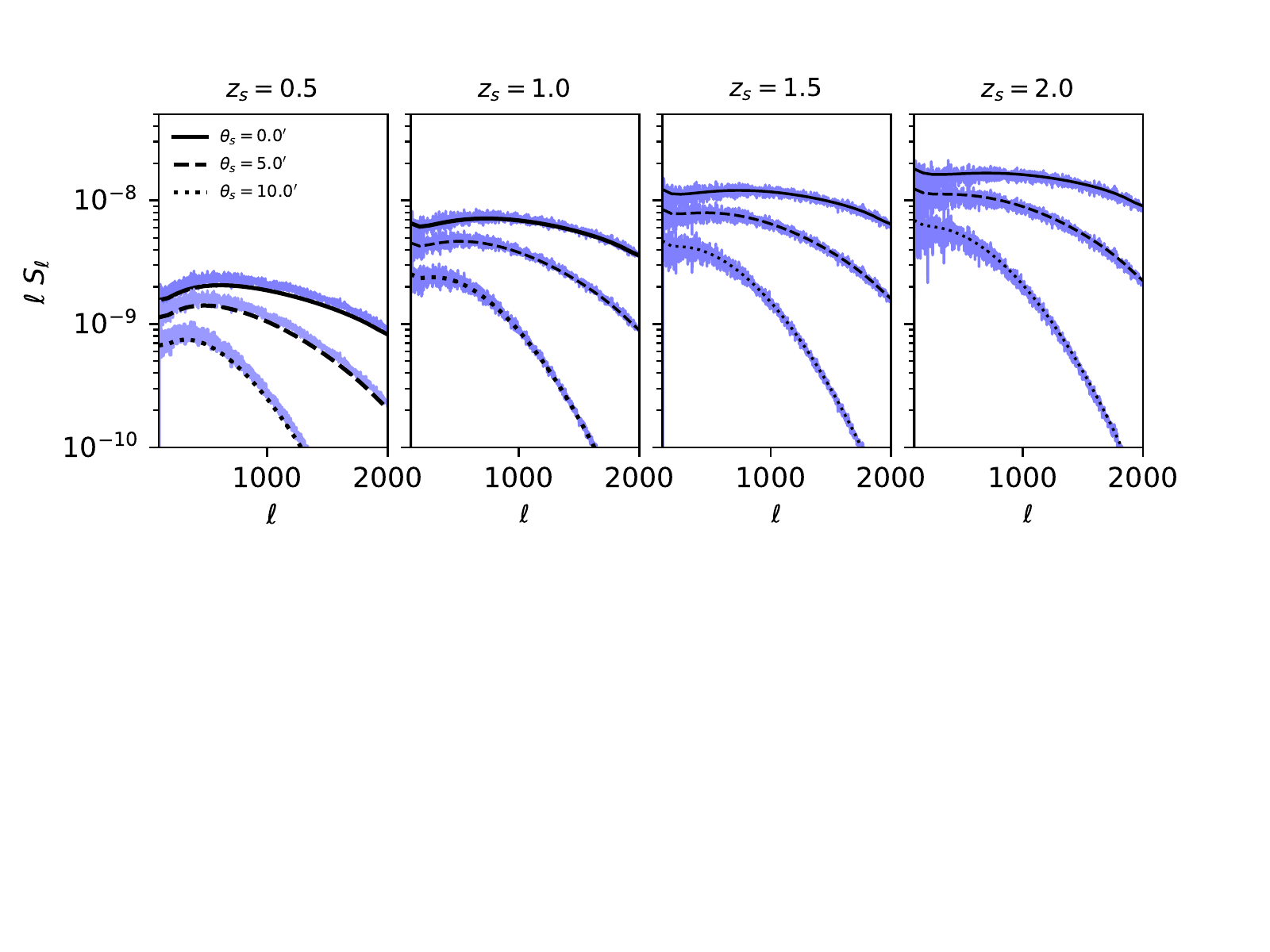}
  \vspace{-5.cm}
  \hspace{1.cm}
  \caption{The skew-spectrum $S_{\ell}$
    defined in Eq.(\ref{eq:define_skewSpec}) is being plotted for various smoothing angles and redshifts.
    The noisy purple lines correspond to results from the simulations.
    The panels from left to right correspond to redshifts $z_s=0.5,1.0,1.5$ and $2.0$.
    In each panel we show skew-spectra corresponding to Gaussian beams with full width at half maxima of $\theta_s=0', 5'$ and $10'$,
    as indicated. We have considered all-sky simulations and no noise was included.
    The theoretical predictions, shown in black, are computed using the expressions Eq.(\ref{eq:defSkew})-Eq.(\ref{eq:defJ}).
    The fitting function of \citep{Taka19} was used throughout this study to model the gravity-induced bispectrum $B_{\delta}$.
    We use all modes below $\ell_{\rm max}= {\rm N}_{\rm side}$ in the computation.}
  \label{fig:allskew}
\end{figure} 
%
Here, we specialize our discussion to the case of weak lensing surveys. The weak lensing convergence $\kappa$
is a line of sight projection of the 3D density contrast $\delta({\bf r})$ at a comoving distance ${\bm r} =(r,{\bm \theta})$ using
a kernel $w(r)$ defined as follows:
\ben
&& \kappa({\bm\theta}) := \int_0^{r_s} dr w(r) \delta(r,{\bm\theta}); \quad
w(r) := {3 \Omega_{\rm M} \over 2} {H_0^2 \over c^2} a^{-1} {d_A(r) d_{A}({r_s-r}) \over d_A(r_s)}.
\een
In this expression, $r=|{\bm r}|$ is the comoving radial distance to the source, $\bm\theta$ describes the angular position on the sky,
$\Omega_M$ is the cosmological matter density parameter (total matter density in units of the critical density),
$H_0$ is the Hubble constant, $c$ is the speed of light, $a = 1/(1+z)$ is the scale factor at a redshift $z$, $d_A(r)$
represents the comoving angular diameter distance at a distance $r$
and $r_s$ is the comoving radial distance to the source plane. The corresponding redshift will
be represented by $z_s$. To keep the analysis simple, in our study we will ignore the source distribution and assume them to be
localized on a single source plane defined by $z_s$. We will study various statistics as a function of $z_s$.
To simplify the analysis we will also ignore photometric redshift errors.
Needless to say, such complications are essential to link predictions to observational data,
and will be presented in an acompanying study.

Fourier decomposing $\delta$ along and perpendicular to the line-of-sight direction we obtain: 
\ben
&& \kappa(\bm\theta) = \int_0^{r_s} dr \omega(r) \int {dk_{\parallel} \over 2\pi}
\int {d^2{\bf k}_{\perp} \over (2\pi)^2} \exp[{\rm i}(rk_{\parallel} + d_A(r)\;{\bm\theta}\cdot{\bf k}_{\perp})]\delta({\bf k}; r).
\label{eq:los}
\een
Here, we have decomposed the 3D wave number ${\bf k}$ along and perpendicular to the radial direction, ${\bf k} = (k_{\parallel}, {\bf k}_{\perp})$
We have used the following convention for the 3D Fourier Transform and its inverse:
\ben
&& \delta({\bf k}) = {1 \over (2\pi)^3}\int d^3{\bf r} \exp(-i {\bf k} \cdot{\bf r}) \delta({\bf r}); \quad
\delta({\bf r}) =  \int d^3{\bf k} \exp(i\,{\bf r}\cdot {\bf k}) \delta({\bf k}).
\een
The corresponding 3D power spectrum and bispectrum for $\delta$ are:
\ben
&& \la\delta({\bk}_{1})\delta({\bk}_{2})\ra_c :=
(2\pi)^{3}\delta_{\rm 3D}({\bk}_{1}+{\bk}_{2})P_{\delta}(k_{1}); \quad k=|{\bf k}|;\\
&& \la\delta({\bk}_{1})\delta({\bk}_{2})\delta({\bk}_{3})\ra_c 
:= (3\pi)^{2}\delta_{\rm 3D}({\bk}_{1}+{\bk}_{2}+{\bk}_{3})B_{\delta}({\bk}_{1},{\bk}_{2},{\bk}_{3}).
\een
Using the {\em small-angle approximation}
the projected power spectrum $P^{\kappa}(l)$ and bispectrum $B^{\kappa}({\bl}_1,{\bl}_2,{\bl}_3)$ of the convergence field $\kappa$ can be expressed
respectively in terms of the 3D power spectrum $P_{\delta}(k)$ and bispectrum  $B_{\delta}({\bf k}_1,{\bf k}_2,{\bf k}_3)$:
\bes
\ben
&& P^{\kappa}({l}) = \int_0^{r_s} dr {\omega^2(r) \over d_A^2(r)}
P_{\delta}\left ({{l} \over d_A(r)}; r \right );\\
&& B^{\kappa}({\bl}_{1},{\bl}_{2},{\bl}_{3}) = \int_0^{r_s} dr {\omega^3(r) \over d_A^4(r)}
B_{\delta}\left ({{\bl}_{1} \over d_A(r)},{{\bl}_{2} \over d_A(r),},
{{\bl}_{3} \over d_A(r)}; r \right).
\een
\ees
Detailed derivations of these expressions can be found in \citep{review}. 
Cross-correlating two-tomographic bins can be used to define cross-spectra $P_{\alpha\beta}^{\kappa}$ and cross-skewspectra $B_{\alpha\beta}^{\kappa}$.
\bes
\ben
&& P_{\alpha\beta}^{\kappa}({l}) = \int_0^{r_{\rm min}} dr {\omega_\alpha(r)\omega_\beta(r) \over d_A^2(r)}
P_{\delta}\left ({{l} \over d_A(r)}; r \right );\\
&& B_{\alpha\beta}^{\kappa}({\bl}_{1},{\bl}_{2},{\bl}_{3}) = \int_0^{r_{\rm min}} dr {\omega_{\alpha}^1(r)\omega_{\beta}^2(r) \over d_A^4(r)}
B_{\delta}\left ({{\bl}_{1} \over d_A(r)},{{\bl}_{2} \over d_A(r),},
{{\bl}_{3} \over d_A(r)}; r \right); \quad r_{min} = min(r_{\alpha},r_{\beta}); \label{eq:tomo_skew_a}\\
&& w_i(r) := {3 \Omega_{\rm M} \over 2} {H_0^2 \over c^2} a^{-1} {d_A(r) d_{A}({r_{si}-r}) \over d_A(r_{si})}; \quad i\in\{\alpha,\beta\}.
\label{eq:tomo_skew_b}
\een
\ees
The integration takes contribution only from the overlapping redshift range of the two bins. Thus, the upper-limit
extends only to the source plane defined by the lower redshift $r_{\rm min} = {\rm min}(r_\alpha,r_\beta)$. Notice that
$P^{\kappa}_{\alpha\beta}=P^{\kappa}_{\beta\alpha}$ but $B^{\kappa}_{\alpha\beta} \ne B^{\kappa}_{\beta\alpha}$ and they carry
independent information. It is possible to directly deal with shear
bispectrum and relate them to density bispectrum thus avoiding the map making process. See \citep{flexions} for bispectra
constructed for higher-spin objects, i.e. shear as well as flexions.
%
\subsection{Skew-spectrum in all-sky and flat-sky}
\label{sec:new_allflat}
%
\begin{figure}
  \begin{center}
    \begin{minipage}[b]{0.99\textwidth}
      \includegraphics[width=0.9\textwidth]{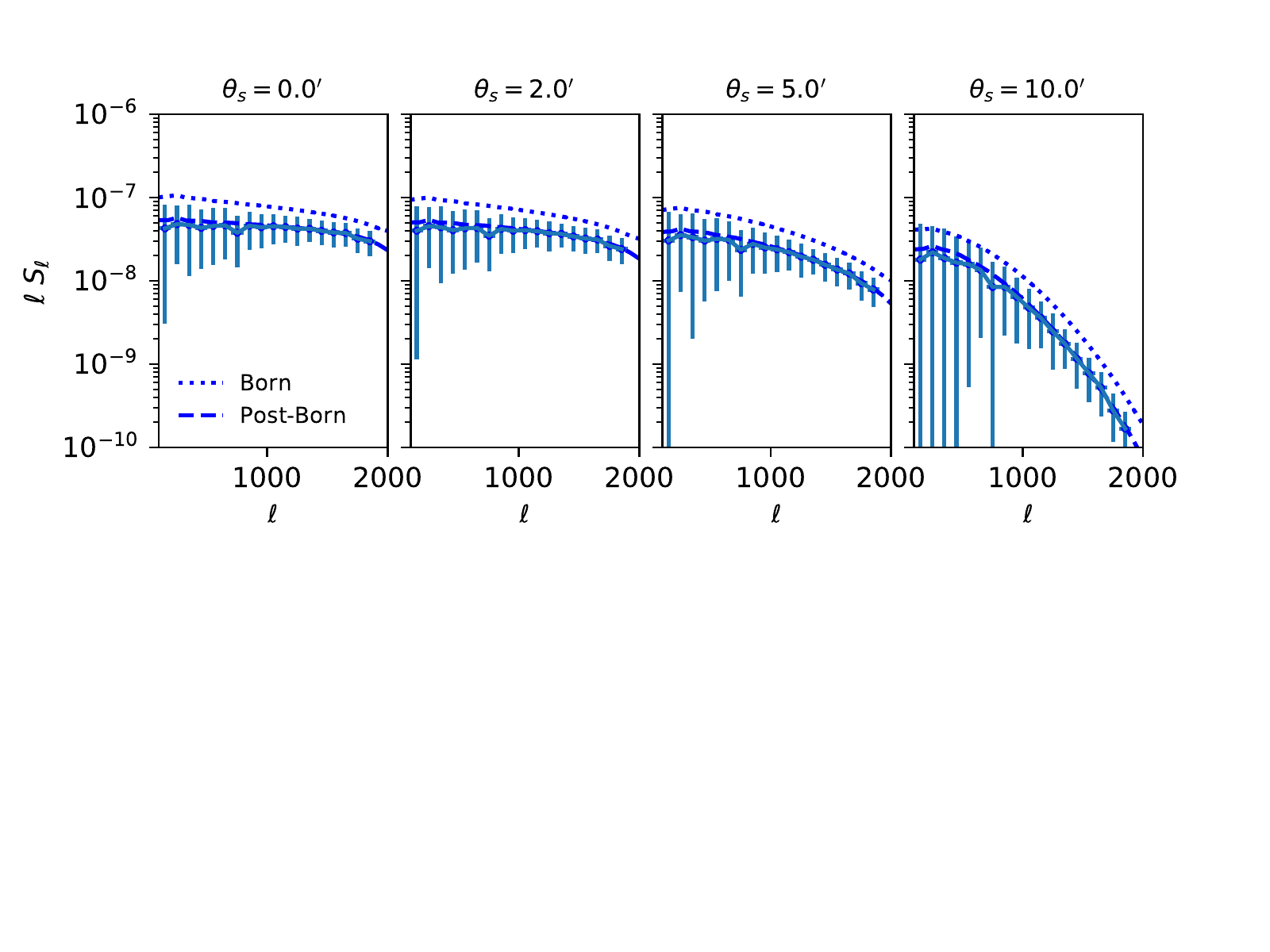}
    \label{fig:cmb}
    \end{minipage}
  \end{center}
  \vspace{-5.cm}
  \caption{The skew-spectra $S_{\ell}$
    defined in Eq.(\ref{eq:define_skewSpec}) for the CMB is being plotted for various smoothing angles.
    The smooth dashed and dotted lines correspond to theoretical predictins.
    The points with error-bars correspond to measurements from numerical simulations.
    The panels from left to right correspond to various smoothing beams of Full Width at Half Maximum (FWHM) $\theta_s=0.',2.0',5.0'$ and $10'$.
    respectively. We have considered all-sky simulations and no noise was included.
    The theoretical predictions are computed using \citep{Taka19}. The dotted curves in each panels
    are computed using the Born approximation, whereas the dashed curves are computed using more accurate post-Born approximation. The error-bars
    are computed using ten different realisations of the simulations. The plots underline the importance of post-Born correction in the computation
    skew-spectrum. Although such corrections can safely be ignored at lower redshifts.}
  \label{fig:cmb}
\end{figure}
%
The skew-spectrum statistic for $\kappa$ is constructed by cross-correlating the
squared $\kappa$ with itself. We start by introducing
the spherical harmonic transform of a convergence map $\kappa(\oh)$
defined over the surface of the sky using spherical harmonics $Y_{\ell m}(\oh)$
to define the multipoles $\kappa_{\ell m}$:
\ben
&& \kappa_{\ell m} := \int\, d{\oh}\, Y_{\ell m}\, \kappa({\oh});
\quad \oh=(\vartheta,\varphi); \;\; d\oh = \sin\vartheta\, d\vartheta\, d\varphi.
\een
Any Gaussian field is completely characterized by its 
power spectrum ${\cal C}^{\kappa}_{\ell}$ which is defined as
${\cal C}^{\kappa}_{\ell}:= \langle \kappa_{\ell m}\kappa^*_{\ell m}\rangle$.
In the flat-sky limit the power spectrum $P^{\kappa}(l)$ is identical
to ${\cal C}^{\kappa}_{\ell}$ at high $\ell$ with the identification $l=\ell$.
The weak lensing $\kappa$ maps are highly non-Gaussian. 
The bispectrum is the lowest-order statistics that characterizes departure from
Gaussianity is defined as the three-point coupling of
harmonic coefficients. The statistics beyond bispectra, e.g.,
the trispectra and its higher-order analogs are increasingly noise dominated.
By assuming isotropy and homogeneity the all-sky bispectrum $B^\kappa_{\ell_1\ell_2\ell_3}$ is defined as:
\ben
&& \la \kappa_{\ell_1 m_1}\kappa_{\ell_2 m_2}\kappa_{\ell_3m_3}\ra_c \equiv 
B^\kappa_{\ell_1\ell_2\ell_3}
\left ( \begin{array} { c c c }
     \ell_1 & \ell_2 & \ell_3 \\
     0 & 0 & 0
\end{array} \right ).
\een
\hspace{4cm}
\begin{figure}
  \begin{center}
  \begin{minipage}[b]{0.99\textwidth}
    \includegraphics[width=0.9\textwidth]{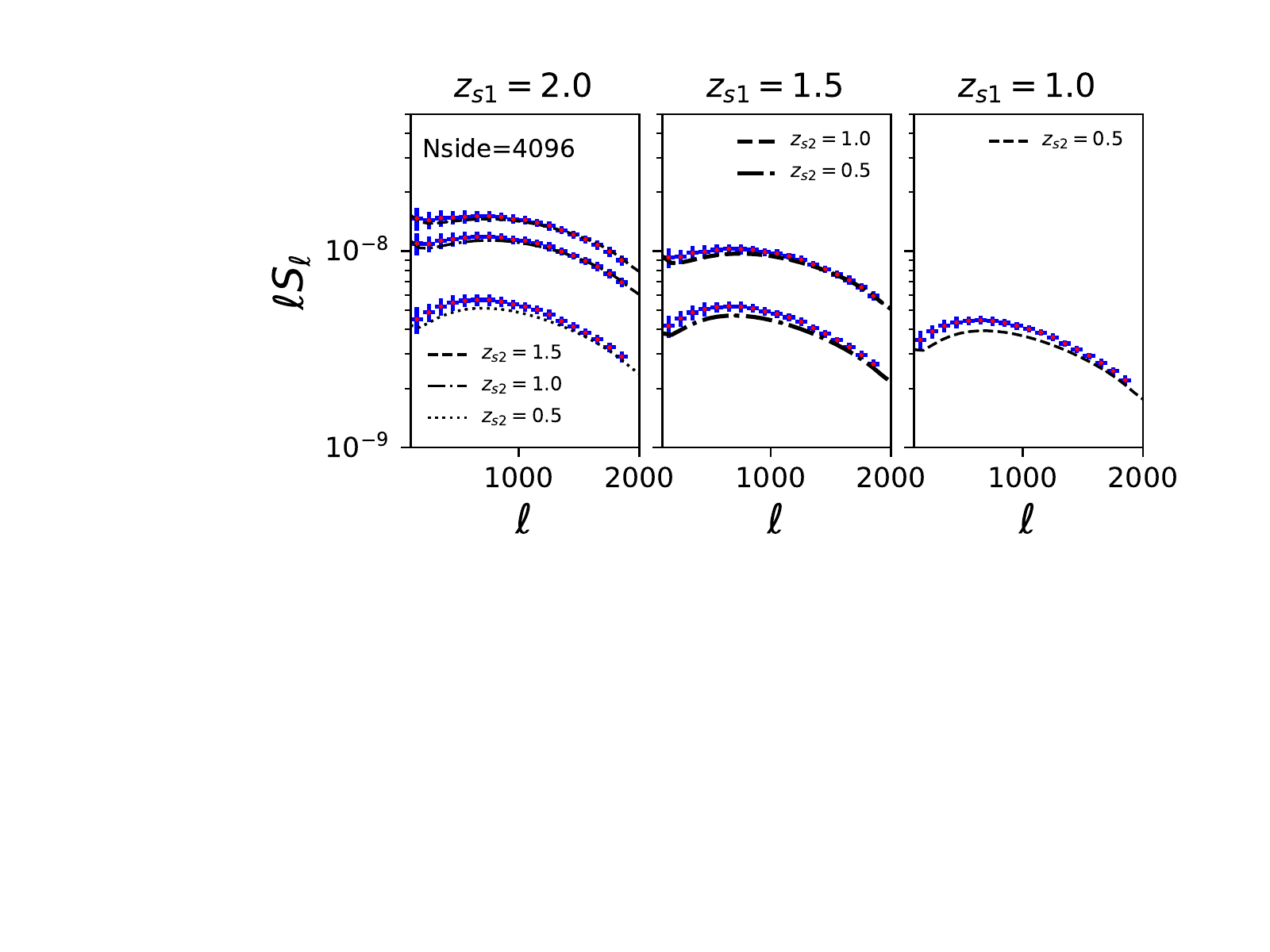}
  \end{minipage}
  \end{center}
  \vspace{-5.cm}
  \caption{The skew-spectra computed by cross-correlating $\kappa^2$ and $\kappa$ from two different redshift bins is plotted.
    In particular, the squared $\kappa_1=\kappa(z_{s1})$ defined for a source redshift $z_{s1}$ and
    $\kappa_{s2}=\kappa(z_{s2})$ at redshift $z_{s2}$ are being cross-correlated
    in the harmonic domain. For this plot we restricted ourselves to $z_{s1}>z_{s2}$.
    The smooth lines correspond to the theoretical predictions and the lines with error bars
    correspond to results from numerical simulations. 
    We use the expression of {\em mixed} bispectrum
    given in Eq.(\ref{eq:tomo_skew_a})-Eq.(\ref{eq:tomo_skew_b}) for computing the theoretical predictions.
    The expression for the estimator for the skew-spectrum is given in Eq.(\ref{eq:define_skewSpec}).
    The panels from left to right
    correspond to $z_{s1}=2.0,1.5$ and $1.0$ respectively and various curves in each panel
    correspond to $z_2$ as indicated. The maps used were constructed at Healpix resolution $N_{\rm side}=4096$. We have filtered all 
    $\ell > 2048$ modes out before analysing them. No additional smoothing was considered. We do not include
    any noise due to intrinsic ellipticity distribution of galaxies. We have used one single all-sky realisation
    to compute the skew-spectra and no mask was included.} 
  \label{fig:cross_skew}
\end{figure}

\noindent
Here the quantity in parentheses is the well-known Wigner-$3j$ symbol which
enforces the rotational invariance. It is only non-zero for the triplets
$(\ell_1,\ell_2,\ell_3)$ that satisfy the {\em trinagular condition} and
$\ell_1+\ell_2+\ell_3$ is even.
The reduced bispectrum $b^\kappa_{\ell_1\ell_2\ell_3}$ is useful in directly
linking the all-sky bispectrum and its flat-sky counterpart.
For the convergence field $\kappa$, $b^\kappa_{\ell_1\ell_2\ell_3}$
is defined through the following expression: 
\ben
&& B^\kappa_{\ell_1\ell_2\ell_3} := \sqrt{(2\ell_1+1)(2\ell_2+1)(2\ell_3+1)\over 4\pi}
\left ( \begin{array} { c c c }
     \ell_1 & \ell_2 & \ell_3 \\
     0 & 0 & 0
\end{array} \right ) b^\kappa_{\ell_1\ell_2\ell_3}.
\een
Finally we are in a position to define the skew spectrum as the cross power-spectra
formed by cross-correlating the squared $\kappa^2(\oh)$ maps against
the original map $\kappa(\oh)$.
\bes
\ben
&& {\cal S}^{}_{\ell} :=
    {1\over 2\ell+1}\sum_m {\rm Real}\{ [\kappa^2]_{\ell m}[\kappa]^*_{\ell m} \}
    =\sum_{\ell_1\ell_2} B^\kappa_{\ell_1 \ell_2\ell} J_{\ell_1\ell_2\ell}; \label{eq:defSkew} \\
&& J_{\ell_1\ell_2\ell} := \sqrt {   {(2\ell_1+1)(2\ell_2+1) \over 4\pi ( 2\ell+1 )}  }
\left ( \begin{array} { c c c }
     \ell_1 & \ell_2 & \ell \\
     0 & 0 & 0
  \end{array} \right ). \label{eq:defJ}
\een
\ees
Here $[\kappa^2]_{\ell m}$ represents the harmonic multipoles computed using a
harmonic decomposition of $\kappa^2$ and $^*$ denotes complex conjugation.
The commonly used (normalised) one-point skewness parameter $S_3={\la\kappa^3\ra_c/\la\kappa^2\ra_c^2}$
can be recovered from the skew-spectrum. The third-order moment $\la\kappa^3\ra$ is given by:
\ben
&& \la\kappa^3(\theta_s)\ra_c =  {1\over 4\pi} \sum_{\ell}(2\ell+1)S^{}_{\ell}{\beta}^3_\ell(\theta_s); 
\een
The smoothing beam (window) is denoted as ${\beta}_\ell(\theta_s)$.
Being a two-point statistic, the skew-spectrum $S^{}_{\ell}$ is related to
the two-to-one correlation function $\xi^{21}$ in the real space. They are related by the
following expression: 
\ben
&& \xi^{21}(\theta_{12}):= \langle\kappa^2(\theta_1)\kappa(\theta_2)\rangle_c =
   {1 \over 4\pi} \sum_{\ell} (2\ell+1) S^{}_{\ell}P_{\ell}(\cos\theta_{12}) {\beta}^3_\ell(\theta_s).
   \een
Here $P_{\ell}$ is the Legendre Polynomial and $\beta_\ell$ is the Gaussian smoothing beam with
Full Width at Half Maximum (FWHM) of $\theta_s$. Suitably normalised two-to-one correlators is 
the lowest order of a family of statistics also known as cumulant correlator
\citep{Bernardeau_bias}, it has also been used in the context of weak-lensing surveys
\citep{Munshi_bias}.
The flat-sky bispectrum is similarly defined through:
\ben
&& \langle \kappa(\bl_1)\kappa(\bl_2)\kappa(\bl_3)\rangle_c =
(2\pi)^2 \delta_{2D}(\bl_1+\bl_2+\bl_3) B^\kappa(\bl_1,\bl_2,\bl_3). 
\een
The flat-sky bispectrum $B^\kappa(\bl_1,\bl_2,\bl_3)$ is identical to the reduced bispectrum
$b_{\ell_1\ell_2\ell_2}$ for high multipole \citep{review_ng}. This can be shown
by using the following asymptotic relationship:
\bes\ben
&& {\cal G}_{\ell_1m_1,\ell_2m_2,\ell_3m_3} \equiv \int d\oh Y_{\ell_1m_1}(\oh)
Y_{\ell_2m_2}(\oh)Y_{\ell_3m_3}(\oh) \nn \\
&&  = \sqrt {   {(2\ell_1+1)(2\ell_2+1)( 2\ell+1 ) \over 4\pi}  }
\left ( \begin{array} { c c c }
     \ell_1 & \ell_2 & \ell \\
     0 & 0 & 0
  \end{array} \right )\left ( \begin{array} { c c c }
     \ell_1 & \ell_2 & \ell \\
     m_1 & m_2 & m_3
\end{array} \right ) \approx (2\pi)^2\delta_{\rm 2D}(\bl_1+\bl_2+\bl_3).
\een\ees
The skew-spectrum in the flat-sky is given by \citep{MunshiPratten}:
\ben
&& \mathcal{S}(l_2) =
\int_0^{\infty} {l_1 d l_1 \over 2 \pi}
\int^{1}_{-1} {d \mu \over 2 \pi \sqrt{1 - \mu^2}}
 B^{\kappa}(\bl_1,\bl_2,-(\bl_1+\bl_2))
\beta(l_1\theta_s)\beta(l_2\theta_s) \beta(| {\bf{l}}_1 + {\bf{l}}_2 |\theta_s).
\label{eq:2ds0} 
  \een 
  In our notation $\mu= ({\bl_1\cdot\bl_2 / l_1 l_2})$, and we have used  $\beta_{\ell_1}(\theta_s) = \beta(l_1\theta_s)$
  to denote the flat-sky beam.
  In the high-$l$ limit we have $\mathcal{S}(l_2) \rightarrow S_{\ell_2}$.

  Here a few comments about the skew-spectrum are in order. The one-point statistics such
  as the skewness parameter has the advantage of having high signal-to-noise.
  However, it lacks distinguishing power as all the available
  information in the bispectrum is compressed into a single number. Therefore,
  such statistics can not distinguish various contributions, e.g.
  from primordial non-Gaussianity or non-Gaussianity from intrinsic alignment of
  source galaxies from the gravity induced secondary non-Gaussianity.
  The skew-spectrum, on the other hand,  retains some of the information regarding the shape
  of the spectrum, thus it can in principle allow to separate
  various contributions or remove possible
  source of contamination from systematics. 

  In this paper we have considered a direct estimator for the skew-spectrum
  as opposed to the optimal estimator developed in \citep{KSW,MunshiHeavens}
  where optimality was achieved by using suitable weights to the harmonics
  that incorporates a match filtering as well as saturates the Cramer-Rao
  limit in the  weakly non-Gaussian limit. Indeed, a simple
  Fisher matrix based analysis will no longer be adequate for
  moderately non-Gaussian weak lensing maps. However optimality
  is not of crucial importance for analysing weak lensing maps as
  the secondary non-Gaussianity is expected to detected with much
  higher signal-to-noise. A direct estimator which is simpler to
  implement will thus be useful for studying non-Gaussianity
  in weak-lensing maps. 

  Next we consider skew-spectrum for specific models of bispectrum considered
  in \textsection\ref{sec:bispec}.
%
%
  \subsection{The skew-spectrum in the tree-level Standard Perturbation Theory (SPT)}
  \label{sec:spt}
%
\hspace{4cm}
\begin{figure}
  \begin{center}
  \begin{minipage}[b]{0.99\textwidth}
    \includegraphics[width=0.9\textwidth]{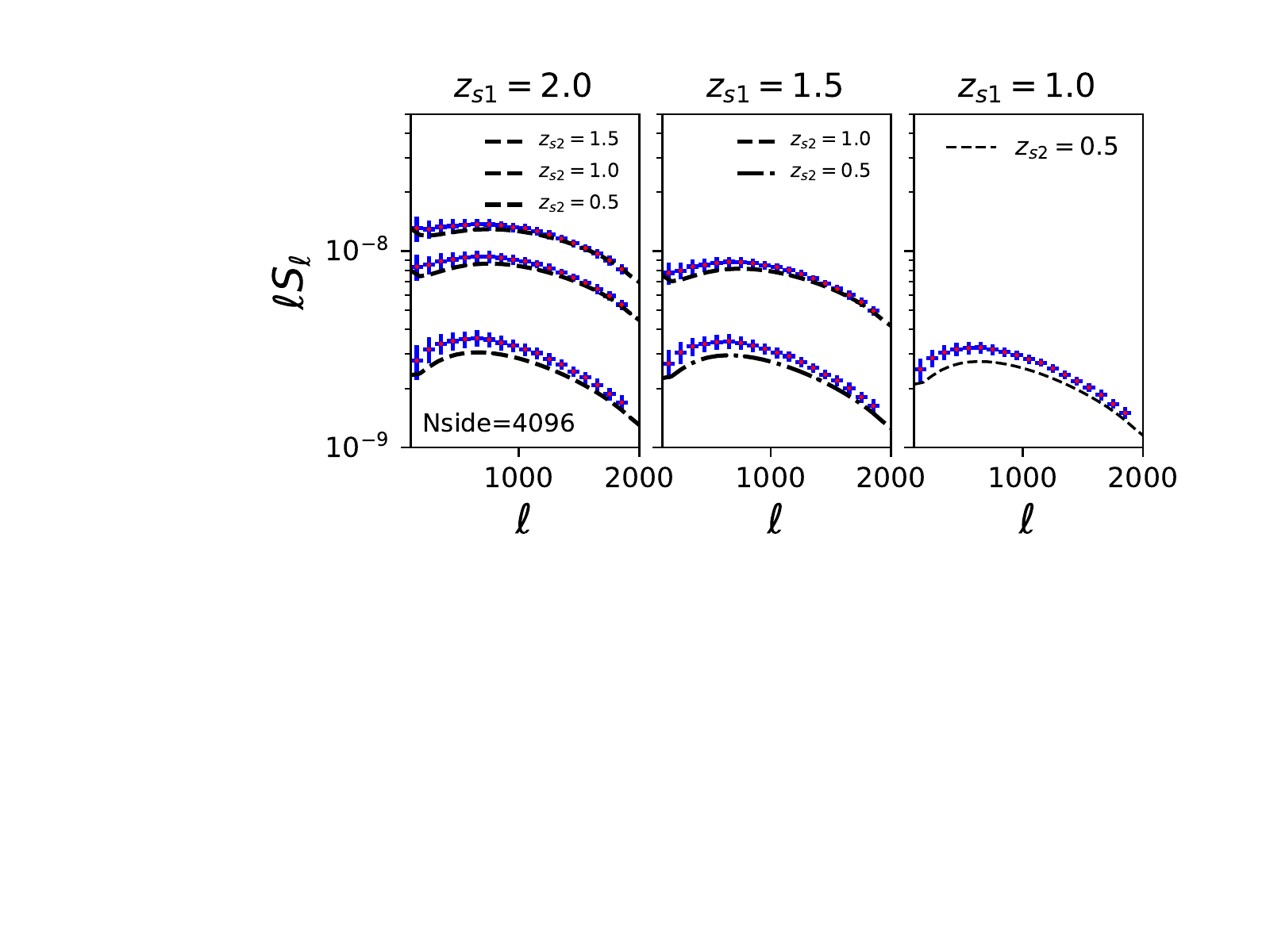}
  \end{minipage}
  \end{center}
  \vspace{-5.cm}
  \caption{Same as Figure-\ref{fig:cross_skew} but the skew-spectrum is being computed cross-correlating $\kappa_1$ and $\kappa_2^2$
    instead of $\kappa_1^2$ and $\kappa_2$ for $z_{s1}>z_{s2}$.}
  \label{fig:cross_skew_spec1}
\end{figure}

\noindent
Analytical predictions for the skew-spectrum for large smoothing angular scales
can be obtained using perturbative calculations.
The low $\ell$ limit of the skew-spectrum and its higher-order
generalisations were recently presented in \citep{MM20}.
This is possible using a technique based on a generating function formalism.
However, for arbitrary $\ell$ an order-by-order calculation is needed.
We will obtain these results using a Gaussian smoothing beam
where complete analytical results in closed form can be derived.
We will consider the gravity induced (secondary) non-Gaussianity.
\bes
\ben
&& S^{}(l_2) =
\int dr\; {w^3(r) \over d_A^4(r)}
\int \frac{l_1 dl_1}{(2 \pi)^2} P_{\delta} \left( {l_1 \over d_a(r)}; r \right)
P_{\delta}\left( {l_2 \over d_a(r)}; r\right) 
\beta^2 (l_1\theta_s) \beta^2(l_2\theta_s) {\cal T}(l_1l_2)  \\
&& {\cal T}(l_1l_2) = \Bigg[  \frac{36}{7} I_0 (l_1 l_2\theta_s^2) - 3 \left( \frac{l_1}{l_2} 
+ \frac{l_2}{l_1} \right) I_1 (l_1 l_2\theta_s^2) + \frac{6}{7} I_2 (l_1l_2\theta_s^2) \Bigg].
\label{eq:F2_SPT}
\een
\ees
\n
The angular integral in Eq.(\ref{eq:2ds0}) can be done analytically using the {\em Modified Bessel Functions}
represented in $I_m$.
To simplify the notation we adopt a parameterisation in terms of the variables $C^{\alpha \beta}_m$:
\bes
\ben
&& S^{}(l_2) =
\,\sigma_L^2 P^{\kappa}(l_2) R_2 \int \frac{l_1 dl_1}{2 \pi} l_1^n  l_2^n \beta^2 (l_1\theta_s) \beta^2(l_2\theta_s) {\cal T}(l_1l_2); \\
&& R_2 = \int_0^{r_s} dr {w^3(r) \over d^{4+2n}_A(r)} D^4_+(z) / \left [ \int_0^{r_s} dr {w^2(r) \over d^{2+n}_A(r)}D^2_+(z) \right ]^2.
\een
\ees
\n
To separate the temporal $r$ and angular $l$ parts of the integral
we replaced the {\em linear} power spectrum $P_{\delta}(k)$ with a power-law form, i.e. $P_{\delta}(k) = A D^2_+(z) k^n$.
Due to the choice of normalisation here the skew-spectrum is independent of the power spectrum amplitude $A$.
The resulting skewness $S_3$ can then be written as:
\ben
&& S_{3} := {\int l_2\, d\,l_2\, } S(l_2)\,
\quad\quad 
\label{eq:define_skewSpec}
\een
The $F_2$ kernel for many modified gravity theories have a similar structural form and can be
treated analytically. Similarly, the Effective Field Theory based approaches introduces
corrective terms to $F_2$ that too have a very similar form \citep{MunshiRegan}. The kernel describing the
primordial non-Gaussianity can also be treated in a similar manner. We will focus on
certain well known cases of modified gravity theories. The analytical results for
these models are important as there are no established numerical fitting-function
available in these scenarios.
\begin{figure}
  \begin{center}
     \label{fig:19}
    \begin{minipage}[b]{0.99\textwidth}
      \includegraphics[width=0.9\textwidth]{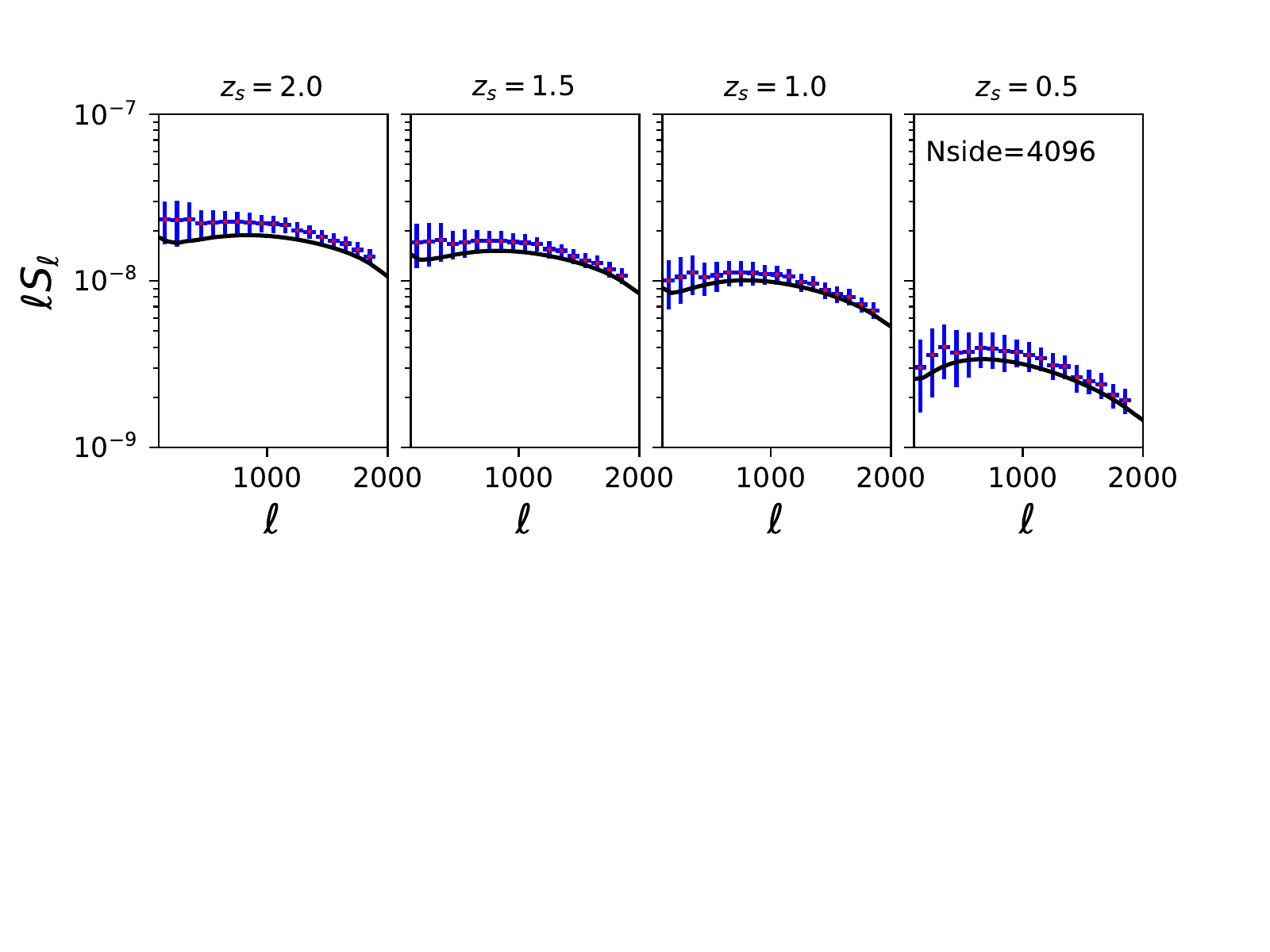}
  \end{minipage}
  \end{center}
  \vspace{-5.cm}
  \caption{The skew-spectra computed by cross-correlating $\kappa^2_{\rm LSS}$ at the last scattering surface of $z_{LSS}=1100$
    and low redshift weak lensing convergence maps at $z_s$ is being plotted.
        As before the solid smooth lines in different panels correspond to theoretical results
    and the lines with error bars correspond to the results from numerical simulations.
    The panels from left to right correspond to
    $z_s=2.0,1.5,1.0$ and $0.5$ respectively. No mask or smoothing was considered. We use the
    expression of {\em mixed} bispectrum
    given in Eq.(\ref{eq:tomo_skew_a})-Eq.(\ref{eq:tomo_skew_b}) for computing the theoretical predictions.}
  \label{fig:cmb1}
\end{figure}  
%
\section{Numerical Simulations}
\label{sec:simu}
%
%
We use the publicly available all-sky weak lensing maps generated by
\citep{Ryuchi}\footnote{http://cosmo.phys.hirosaki-u.ac.jp/takahasi/allsky\_raytracing/}
that were generated using ray-tracing through N-body simulations.
The underlying simulations followed the gravitational clustering of $2048^3$ particles.
Multiple lens planes were used to generate convergence $\kappa$ and the corresponding shear $\gamma$ maps.
To generate the maps in these simulations, the source redshifts used were in the range $z_s= 0.05-5.30$ at an interval $\Delta z_s = 0.05$.
In our study, we have used the maps with $z_s=0.5,1.0,1.5, 2.0$.
For CMB maps the lensing potentials were constructed using the deflection angles
which were used to construct the lensing potentials and eventually the $\kappa$ maps.
In recent studies inclusion of post-Born terms in lensing statistics were outlined \citep{LewisPratten}.
The maps we use include post-Born corrections.
In this study we will see that at the low source redshift such corrections play a negligible role although they play a significant role
at higher redshift, e.g. in case of lensing of CMB.
The convergence maps were generated using an equal area pixelisation scheme
in {\tt HEALPix}\footnote{https://healpix.jpl.nasa.gov/} format\citep{Gorski}.
In this pixelisation scheme the number of pixels scale as $N_{\rm pix} = 12 N^2_{\rm side}$
where $N_{\rm side}$ is the resolution parameter which can take values $N_{\rm side} = 2^{N}$ with $N=1,2,\cdots$.
The set of maps we use in this study are generated at $N_{\rm side}=4096$ and were cross-checked against higher resolution maps
constructed at a resolution $N_{\rm side}=8192, 16384$ for consistency. These maps constructed at different resolution were found to be
consistent with each other up to the angular harmonics $\ell \le 2000$.
Various additional tests were also performed using an Electric/Magnetic (E/B) decomposition of the shear maps
for the construction of $\kappa$ maps \citep{Ryuchi}.
We have used high resolution maps $N_{\rm side} =4096$.
We have degraded these maps to $N_{\rm side} =2048$ and analysed them
for harmonic modes satisfying $\ell < 2 N_{\rm side}$. 
The background cosmological parameters used for these simulations are: $\Omega_{\rm CDM} = 0.233$, $\Omega_b = 0.046$,
$\Omega_{\rm M} = \Omega_{\rm CDM}+\Omega_b, \Omega_{\Lambda}=1-\Omega_{\rm M}$ and $h=0.7$.
The amplitude of density fluctuation $\sigma_8=0.82$ and the spectral index $n_s=0.97$.
Examples of $\kappa$ maps used in our study are presented in  Figure-\ref{fig:allsky}.
It is also worth mentioning that these maps were also used to recently analyze the bispectrum
the context of CMB lensing \citep{Namikawa18} and for weak lensing of galaxies at low redhifts \citep{MunshiNamikawa19,MunshiMcEwen20}.
%
%
%
%
\begin{figure}
  \begin{center}
    \begin{minipage}[b]{0.99\textwidth}
      \includegraphics[width=0.9\textwidth]{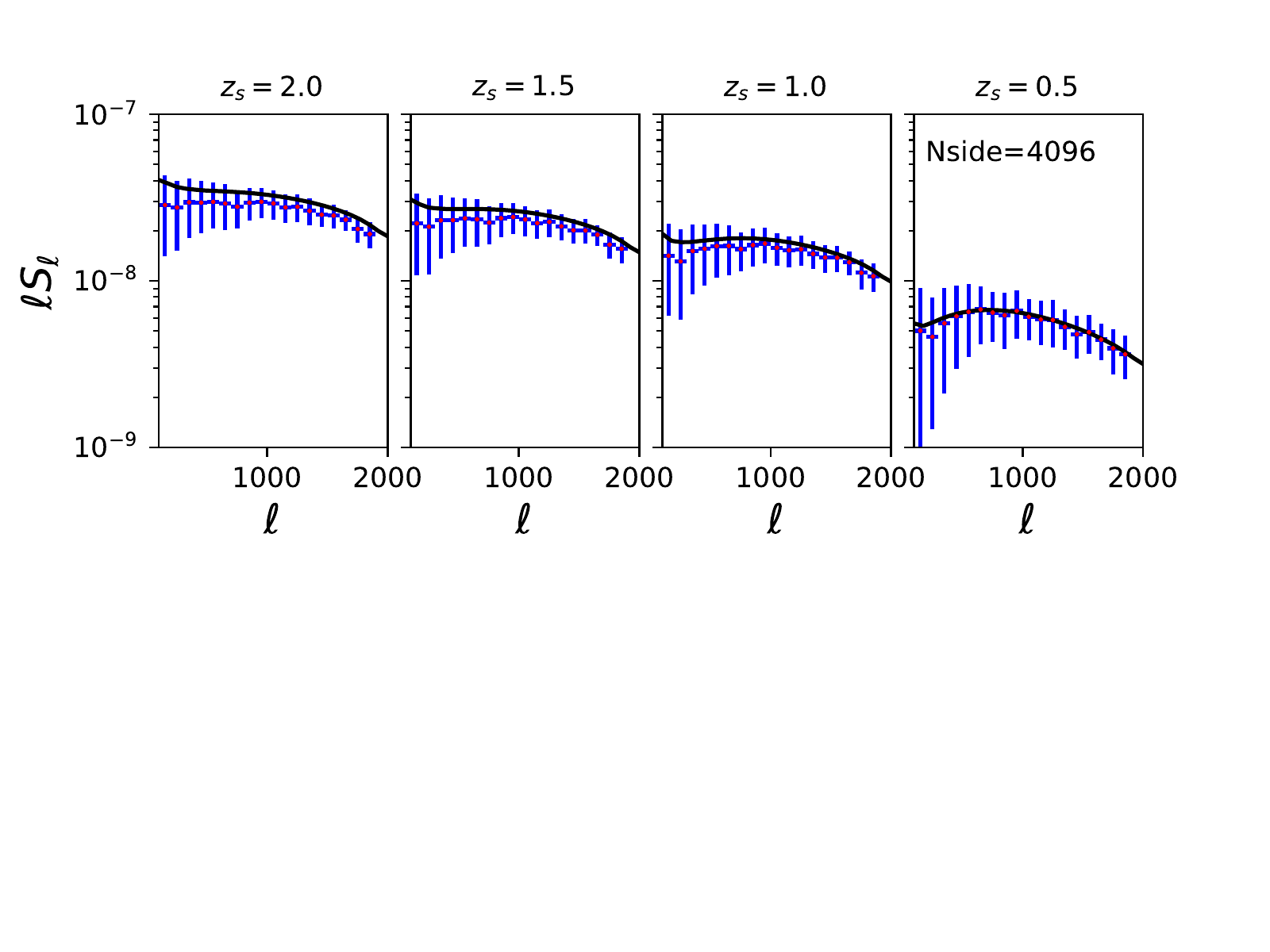}
    \label{fig:19}
  \end{minipage}
  \end{center}
  \vspace{-5.cm}
  \caption{Same as Figure-\ref{fig:cmb1} but are constructed by cross-correlating $\kappa_{\rm LSS}$ against $\kappa^2_{s}$.}
  \label{fig:cmb_cross_p}
\end{figure}
%
\section{The Pseudo Skew-Spectrum Estimator}
\label{estimator}
%
The optimal Maximum Likelihood (ML) estimators or the quadratic maximum likelihood (QML) estimators \citep{GPE} are often
used for analyzing cosmological data sets. The optimality of these estimators
require inverse covariance weighting of the input data vector which clearly is not practical
for large cosmological data sets though various clever algorithmic techniques have been considered \citep{SPO}.
As a result many sub-optimal estimators, which use heuristic  weighting schemes, have been developed.
The so-called pseudo-$\cal C_{\ell}$ (PCL) technique was introduced in \citep{Hivon_Master} in the harmonic domain.
Later a related correlation function based approach was introduced in \cite{Spice}.
These estimators are unbiased but are sub-optimal. Typically various heuristic
weighting depending on sky coverage, as well as noise characteristics can improve
the optimality of these estimators typically in noise dominated high-$\ell$ (or smaller angular scales) regime.
The maximum likelihood estimators on the other hand can be efficiently used for larger
smoothing scales. Different hybridization schemes can be used to combine the large angular scale
(equivalently the low $\ell$) estimates
using QML with small angular scale (high $\ell$) PCL estimates \citep{GPE}.
In our study we will use a direct pseudo-$\cal C_{\ell}$ estimator for the skew-spectrum.
The direct estimator from the masked sky ${\tilde S}_{\ell}$ is related  
to the underlying all-sky $S_{\ell}$ skew-spectrum through a mode-mixing matrix $M_{\ell\ell'}$
that depends on the mask. 
\bes
\ben
&& {\tilde S}_{\ell} = M_{\ell\ell^{\prime}} S_{\ell} \label{eq:tilde}; \quad
{\hat S}_{\ell} = M^{-1}_{\ell\ell^{\prime}}{\tilde S}_{\ell} \label{eq:hat}; \quad \langle {\hat S}_{\ell}\rangle  = S_{\ell}.
\label{eq:PCL1}
\een
\ees
Here $\tilde{\cal S}_{\ell'}^{(21)}$ denotes the skew-spectrum computed from a map in the presence of a mask $w(\oh)$,
$\hat{\cal S}_{\ell'}^{(21)}$ is the all-sky estimate.
The mode-coupling matrix $M_{\ell\ell'}$ is given in terms of the power spectrum of the mask $w(\oh)$ as follows:
\ben
&& M_{\ell\ell'} = (2\ell'+1)\sum_{\ell''}
\left ( \begin{array}{ c c c }
     \ell & \ell' & \ell'' \\
     0 & 0 & 0
  \end{array} \right)^2
{ (2\ell'' +1 )\over 4\pi} |w_{\ell''}^2|; \label{eq:MLL1}
\een
Here $w_{\ell} = {1/(2\ell+1)} \sum_m w_{\ell m} w^*_{\ell m}$ is the
power spectrum of the mask constructed from the harmonic-coefficient $w_{\ell m}$ of the map.
The {\em coupling} matrix $M_{\ell\ell^{\prime}}$
encodes the mode-mixing due to the presence of a mask.
We have used this estimator for estimation of skew-spectrum from individual tomographic bins
as well as cross-correlating two different bins. In case of cross-correlation we have
used the same mask for the two different bins.
The generalization of the PCL method to estimate higher-order spectra were developed in \citep{xn1,xn2} 
for spin-0 fields and in higher spin fields in \citep{flexions} as well as in 3D in \citep{MunshiKitching}.

In our study, we have used the mask which is shown in Figure-\ref{fig:myhealpymask}.
To construct this mask all pixels (shown in maroon) lying within $22$ degree of either
the galactic or ecliptic planes are discarded. The remaining unmasked pixels cover $14,490$ degree$^2$
of the sky, making fraction of the sky covered $f_{\rm sky} \approx 0.35$ \citep{Taylor}.
Various aspects of noise involved in cross-correlating CMB lensing maps and galaxy lensing maps
are discussed in \citep{Fabbian19}.

Typically, to construct an unbiased PCL estimator the noise contribution is subtracted from the
total estimates. This however is not necessary for the construction of the skew-spectrum estimator as the bispectrum of a Gaussian noise
is zero. However, presence of noise in the data does increase the
variance of the estimator. We will not attempt to construct the covariance matrix of our estimator.
Such a generalization will be presented in a future publication.
%
%
\section{Results and Discussion}
\label{sec:disc}
%
%
 In this section we discuss the numerical results presented in this paper.
  We have used all-sky simulations generated at $N_{\rm side}=4096$ for validating the theoretical
  predictions. We have used the simulations generated at lower redshifts for weak lensing
  studies along with the lensing maps generated at $z_s=1100$ (last scattering surface).
  Examples of the maps used in our study are presented in Figure-\ref{fig:allsky}.
  The mask used in studying the effect of mask on estimation of skew-spectrum
  is presented in Figure-\ref{fig:myhealpymask}.
  We have used these maps for constructing the skew-spectra at individual redshift
  as well as computing the skew-spectrum by cross-correlating two different redshifts.
  Below we list our findings.
\begin{figure}
  \begin{center}
    \begin{minipage}[b]{0.3\textwidth}
      \includegraphics[width=\textwidth]{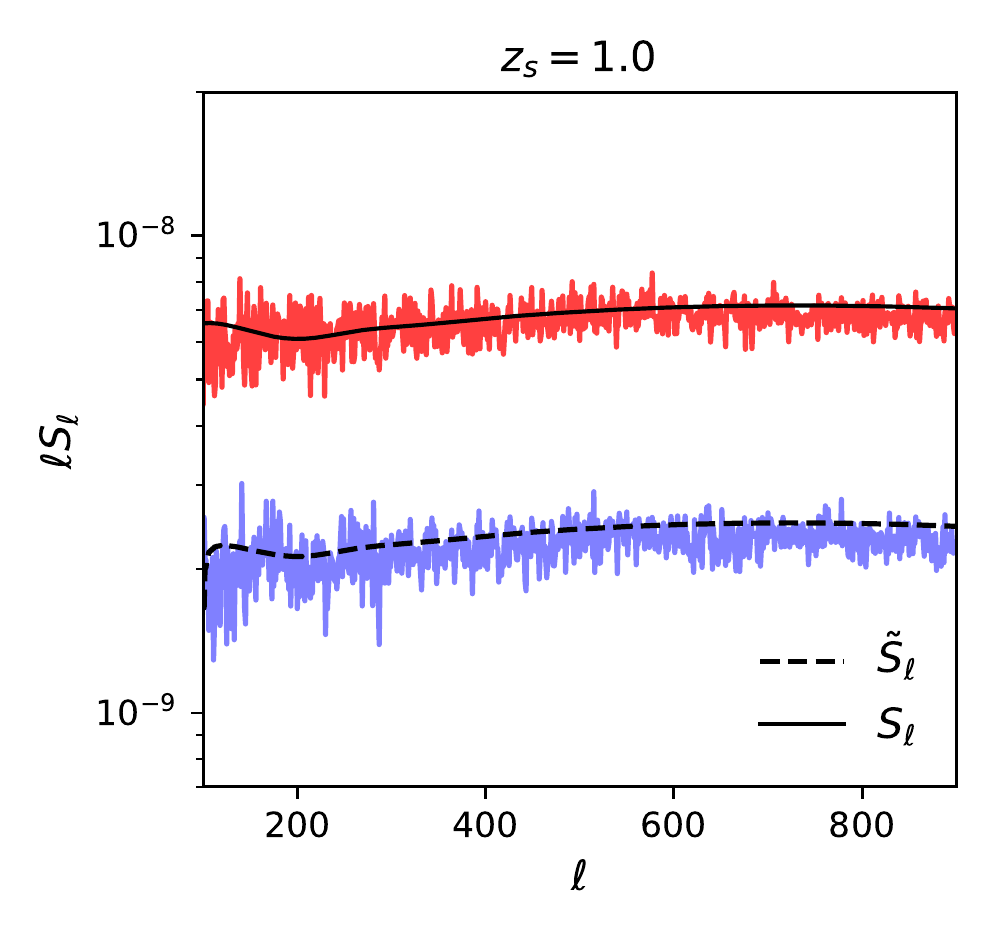}
    \end{minipage}
    \begin{minipage}[b]{0.3\textwidth}
      \includegraphics[width=\textwidth]{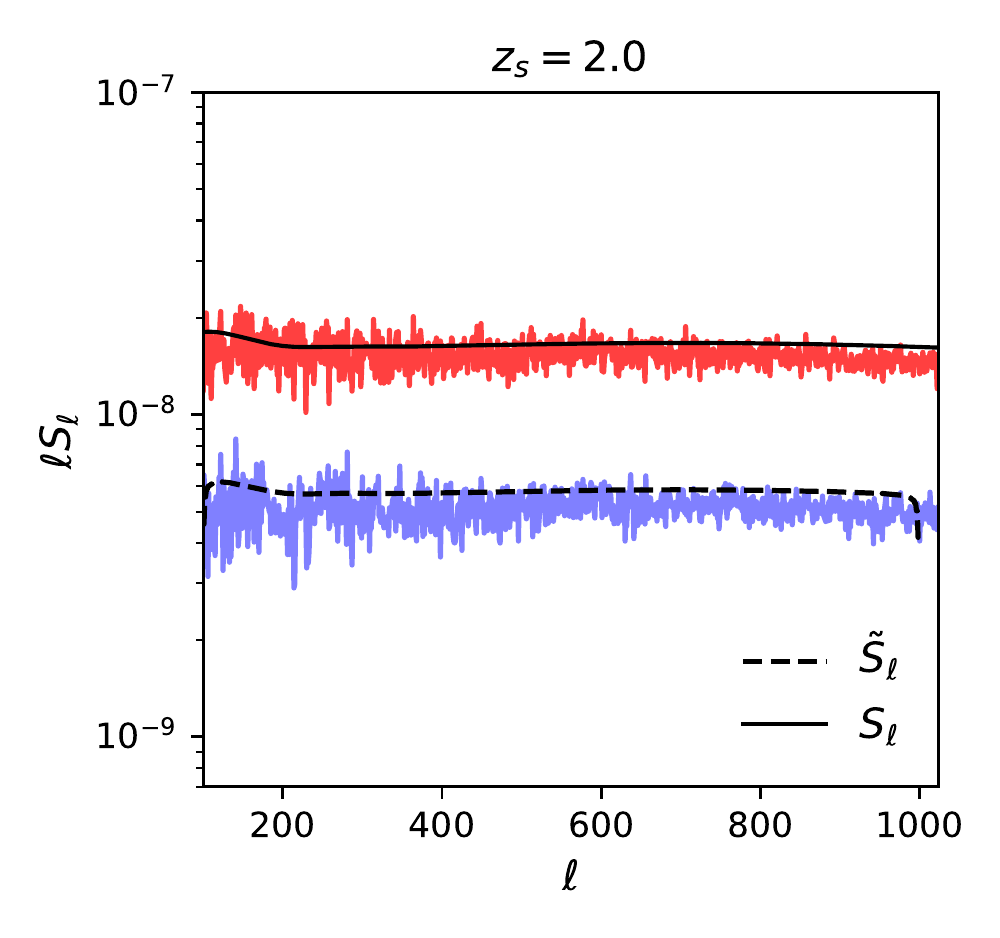}
  \end{minipage}
  \end{center}
  \vspace{0.25cm}
  \caption{The pseudo-${\cal S}_{\ell}$ for two different redshifts are presented.
    A Euclid type mask was used in our study. The regions which are within $22\, \rm deg$\, of
    the galactic or ecliptic plane are removed from our study. The fraction of the sky left unmasked is $f_{\rm sky} = 0.35\;$
    (roughly $14,490$\,deg$^2$ of the sky). The left (right) panel corresponds to $z_s=1.0 (z_s=2.0)$. The upper smooth solid curves
    represents the theoretical $S_{\ell}$. The two upper curves represent estimated skew-spectrum from a single realisation.
    The curve that shows more scatter represents skew-spectrum estimated from an {\em all-sky} map. The curve whcih shows
    more scatter correspond to ${\hat S}_{\ell}$ using Eq.(\ref{eq:hat}) and Eq.(\ref{eq:tilde}).}
  \label{fig:mask}
\end{figure} 

\begin{figure}
  \begin{center}
    \begin{minipage}[b]{0.3\textwidth}
      \includegraphics[width=\textwidth]{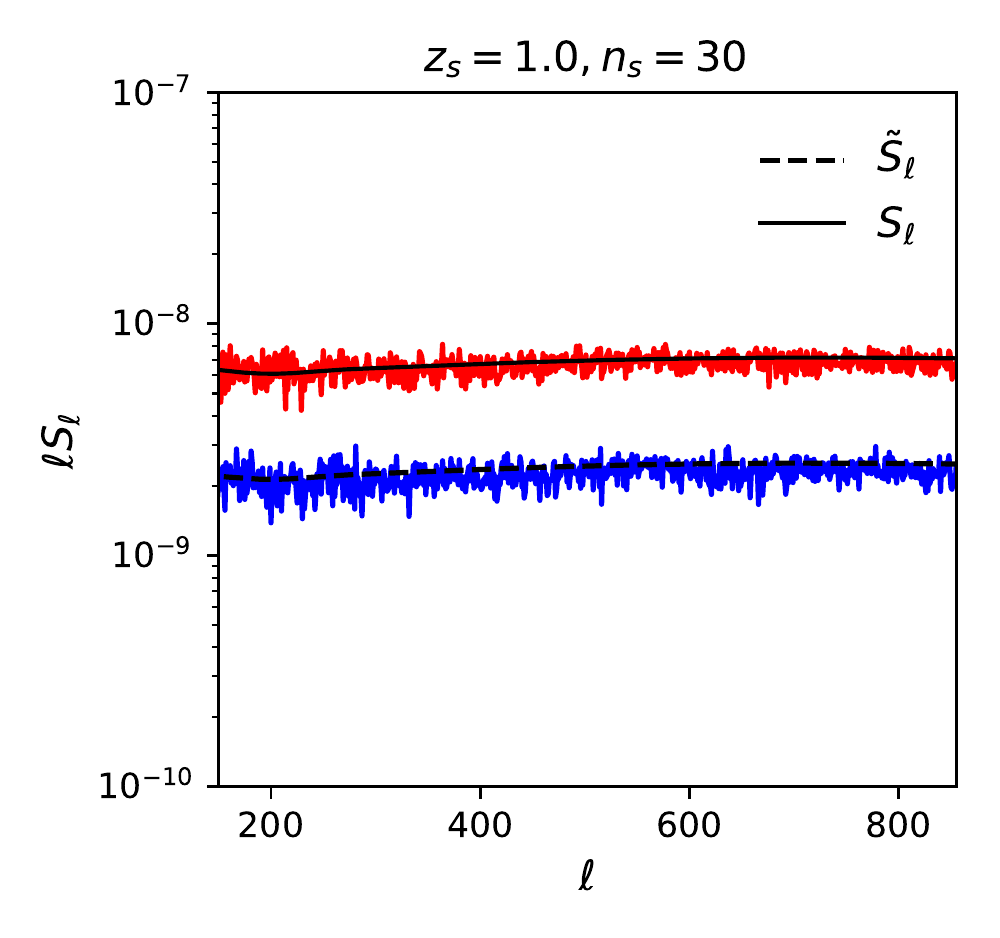}
    \end{minipage}
    \begin{minipage}[b]{0.3\textwidth}
      \includegraphics[width=\textwidth]{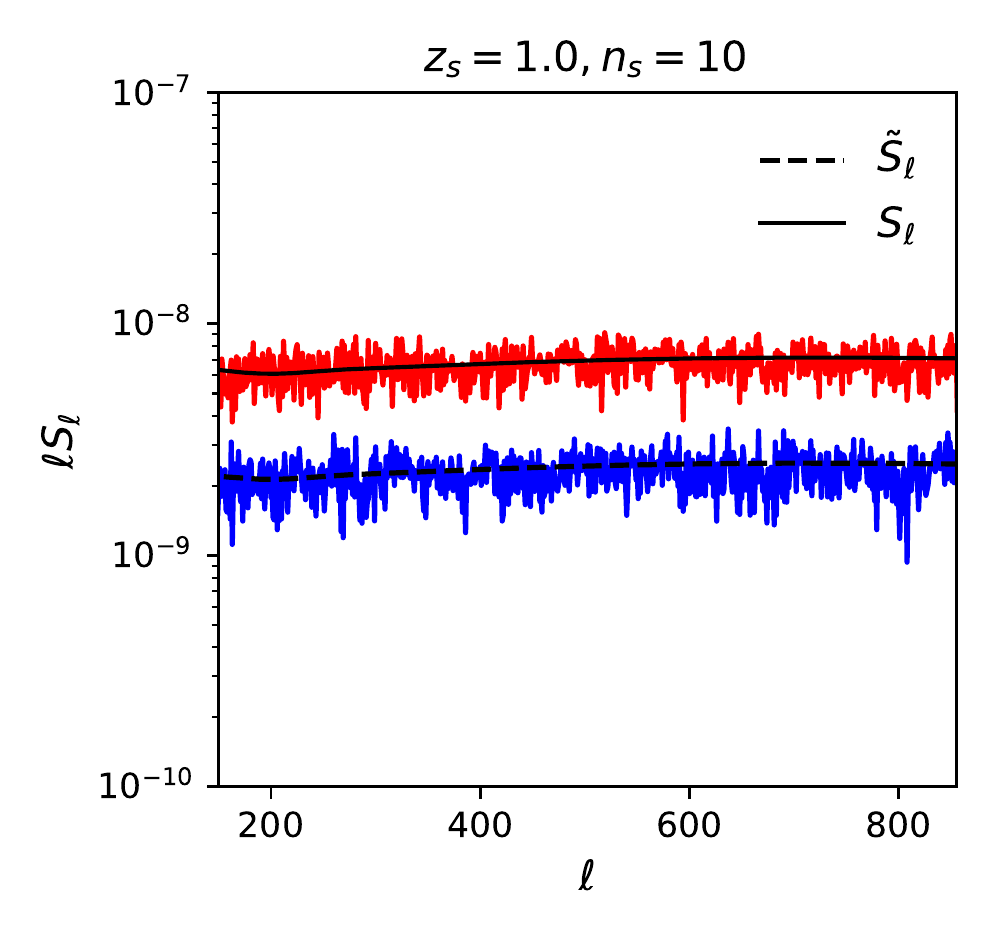}
  \end{minipage}
  \end{center}
  \vspace{0.25cm}
  \caption{The impact of noise (assumed Gaussian) on estimation of skew-spectrum is presented. In both panels
    the source plane is fixed at $z_s$=1. The solid lines in each panel represent the theoretical skew-spectrum
    for $z_s=1$. The dashed line represent the pseudo skew-spectrum or $\tilde S_l$ for the Euclid-type mask
    being considered. Inclusion of Gaussian noise increases the scatter but the estimator remains unbiased.
    The left-panel corresponds to source density of $n_s=30\; {\rm arcmin}^{-2}$ and the right-panel corresponds to $n_s=10\; {\rm arcmin}^{-2}$.}
  \label{fig:noise}
\end{figure}

\begin{figure}
    \begin{minipage}[b]{0.33\textwidth}  
      \includegraphics[width=\textwidth]{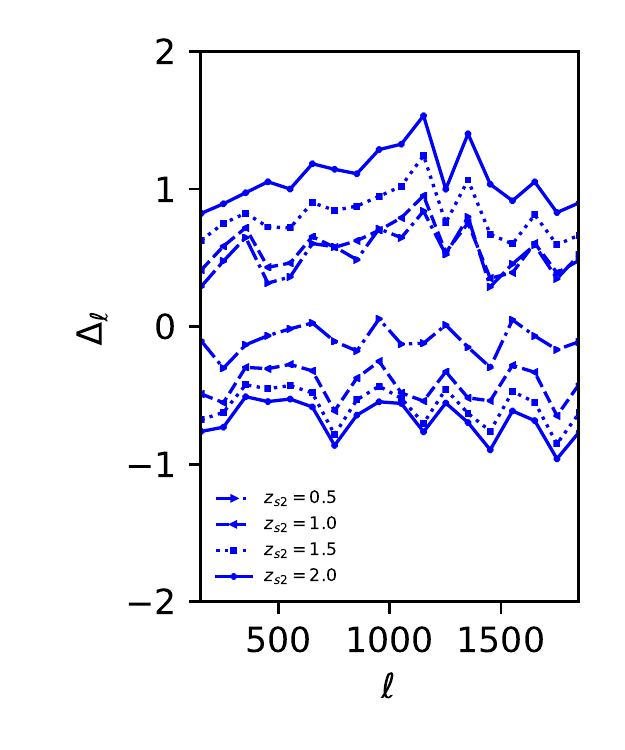}
    \end{minipage}
  \begin{minipage}[b]{0.33\textwidth}
    \includegraphics[width=\textwidth]{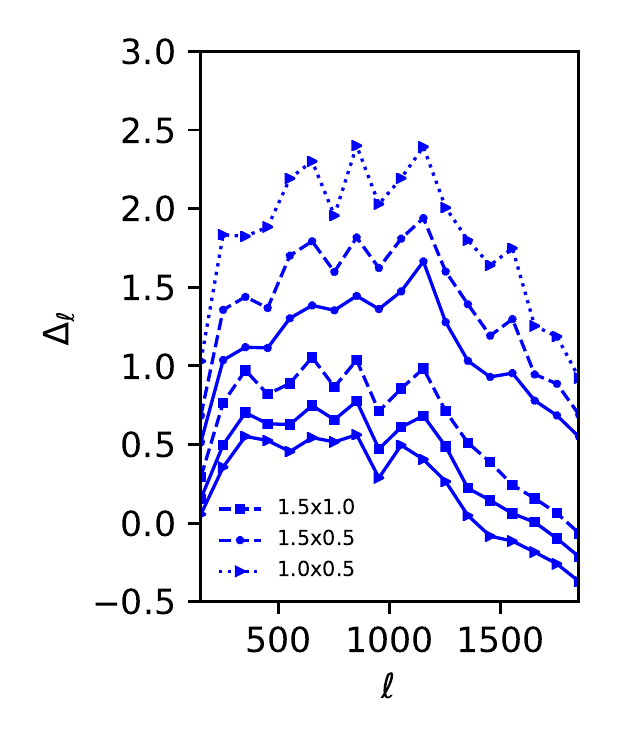}
  \end{minipage}
  \begin{minipage}[b]{0.33\textwidth}
    \includegraphics[width=\textwidth]{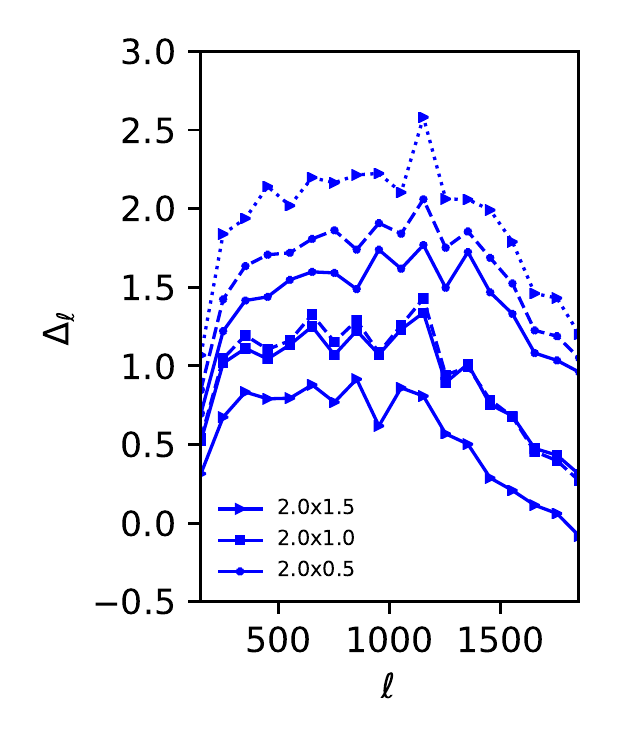}
  \end{minipage}
  \caption{In this figure we present the accuracy of the fitting function we have used in our study.
    We have used the binned skew-spectra for our comparision. A bin-size of $\delta_b = 100$ is being used. 
    Here $\Delta_{\ell}$ represents the {\em normalised} deviation from predictions and results from simulations (see text for details).
    The left-panel corresponds to the skew-spectrum computed by cross-correlating the CMB sky and the low redshift weak lensing
    The middle- and the right-panel correspond to the skew-spectra obtained by cross-correlating two tomographic lensing maps.
    The plots in the left-panel are obtained using the results presented in the Figure-\ref{fig:cmb1} and Figure-\ref{fig:cmb_cross_p}.
    The upper set of curves correspond to the cross-correlation of tomographic bins against convergence map at $z_{\rm LSS}$ i.e.
    $\langle \kappa^2(z_{s})\kappa_{\rm LSS}\rangle$,
    whereas the lower curves correspond to their symmetric counterparts $\langle \kappa(z_{s})\kappa^2_{\rm LSS}\rangle$.
    The error in skew-spectra computed using lower redshift maps that are  associated
    with $\langle\kappa^2(z)\kappa_{\rm LSS}\rangle$ are depicted in the middel panel.
    Their symmetric counterparts are shown in the right-panel. The line-styles used in the middle and right panels are identical.
    For the corresponding skew-spectra see Figure-\ref{fig:cross_skew} and Figure-\ref{fig:cross_skew_spec1} respectively.}
\label{fig:testerror}
\end{figure}

\begin{enumerate}
\item
  {\bf Skew-spectra from Individual Tomographic Bins:}
  First, we compute the theoretical skew-spectra $S_{\ell}$ using Eq.(\ref{eq:define_skewSpec}) as a function of harmonics $\ell$
  for various smoothing angular scales as well as redshifts.
  The results for lower redshift bins are plotted in Figure -\ref{fig:allskew} and the results for the last scattering surface (LSS)
  is plotted in Figure -\ref{fig:cmb}.
  In Figure -\ref{fig:allskew} the panels from left to right correspond to redshifts $z_s=0.5,1.0,1.5$ and $2.0$.
  In each panel three different smoothing angular scales are considered, from top to bottom
  the curves correspond to full width half maxima (FWHM) of the Gaussian beam $\theta_s=0.0',5.0'$ and $10'$ as indicated.
  For the CMB sky shown in Figure -\ref{fig:cmb} the panels from left to right correspond to
  four different Gaussian beams $\theta_s=0.0',5.0'$ and $10'$. We have computed the
  skew-spectra using the Born-approximation as well as including the post Born correction terms \citep{LewisPratten}.
  We found that the post-Born corrections will be important in modelling the skew-spectra at high redshifts.
  However, for the lower redshifts we found this corrections to be negligible as expected. 
  We have considered all-sky simulations and no noise was included.
  The theoretical predictions are computed using the expressions Eq.(\ref{eq:defSkew})-Eq.(\ref{eq:defJ}).
  The fitting function of \citep{Taka19} was used throughout in this study to model the gravity-induced bispectrum $B_{\delta}$.
  We use all modes below $\ell_{max}= 2{\rm N}_{\rm side}$ in our computation. We have also removed all modes
  $\ell_{max} < 100$ from our computation. These fitting-functions are found to
  be an excellent description of the simulated data.
\item
  {\bf Cross-correlating Two Tomographic Bins:}
  The skew-spectra computed by cross-correlating $\kappa^2(\oh)$ and $\kappa(\oh)$ from two different redshift bins is being plotted
  in Figure-\ref{fig:cross_skew} and Figure-\ref{fig:cross_skew_spec1}.
  In particular, squared $\kappa_1=\kappa(z_1)$ defined for a source redshift $z_1$ and
  $\kappa_2=\kappa(z_2)$ at redshift $z_2$ is being cross-correlated
  in the harmonic domain. For this plot we restrict ourselves to $z_1>z_2$.
  We use the expression of {\em mixed} bispectrum
  given in Eq.(\ref{eq:tomo_skew_a})-Eq.(\ref{eq:tomo_skew_b}) for computing the theoretical predictions.
  The expression for the estimator for the skew-spectrum is given in Eq.(\ref{eq:define_skewSpec}).
  From left to right
  panels correspond to $z_1=2.0,1.5$ and $1.0$ respectively and various curves in each panel
  correspond to $z_2$ as indicated. The maps used were constructed at Healpix resolution $N_{\rm side}=4096$. We have filtered all 
  $\ell > 2048$ modes out before analyzing them. No additional smoothing was considered. We do not include
  any noise due to intrinsic ellipticity distribution of galaxies. We have used one single all-sky realization
  to compute the skew-spectra and no mask was included.
  The skew-spectra constructed by cross-correlating $\kappa^2$ at the Last Scattering Surface (LSS, $z_s=1100$) and $\kappa$ at lower redshift
  is presented in Figure-\ref{fig:cmb1}. Similarly, the skew-spectrum constructed using $\kappa$ at LSS and
  $\kappa^2$ at lower redshift is presented in Figure-\ref{fig:cmb_cross_p}. We found that the post-Born correction is
  negligible in modelling the skew-spectrum constructed cross-correlating maps from two redshifts.

\item
{\bf Accuracy of Predictions:}
  To quantify the difference of predicted skew-spectra and the one estimated from numerical simulation we have
  used the following statistics:
  \ben
  && \Delta_b = {1 \over \sigma_b}\left [ \hat{S_b} - S^{th}_b \right ].
  \een
  {Here $\hat S_b$ represents the binned theoretical skew-spectrum and $S^{th}_b$ is the estimated binned skew-spectrum
  from numerical simulation and $\sigma_b$ is the standard deviation of the fluctuations in individual $\ell$ modes
  within a bin. We have chosen a bin-size of $\delta_b$. The results are shown in Figure-\ref{fig:testerror}.
  The left panel shows the errors in skew-spectra obtained by cross-correlating $\kappa^2_{\rm LSS}$ and low redshift
  $\kappa$ (upper curves) and their symmetric counterparts (lower curves). The fitting functions under-predict
  the simulation results for $\langle \kappa^2_{\rm LSS}\kappa\rangle$ and under-predicts
  the results for skew-spectra associated with $\langle \kappa_{\rm LSS}\kappa^2\rangle$. The difference
  between theory and simulation is lowest for $z_s=0.5$ and increases with the redshift. For $z_s=2.0$ it can
  be as high as $1.5\sigma_b$. The results are more pronounced for the intermediate bins.
  The middle- and right panels of Figure-\ref{fig:testerror} depicts $\Delta_b$ for $\kappa_1^2\kappa_2$ (middle-panel)
  and $\kappa_2^2\kappa_1$ respectively. The difference is highest for skew-spectra involving $z_s=0.5$ and
  lower for higher redshift $z_s=2.0$. The $\Delta_b$ can reach a value of $2.5$ for lower redshifts.
  The theory typically under-predicts the data.}
  
  {The individual skew-spectral bins are correlated as the skew-spectrum is an integrated measure, i.e., individual $\ell$ modes (bins)
  depend on the entire range of $\ell$ modes (bins). So a straight forward $\chi^2$ analysis (using a diagonal covariance matrix) is not possible.
  Nevertheless, notice that we have considered noise-free simulations in charachterization of errors.
  Inclusions of noise will increase $\sigma_b$ and decrease $\Delta_b$.
  We have considered full-sky maps but, inclusion of the masks will increase the scatter and thus further reduce the
  value of $\Delta_b$. Hence, the deviations seen here should be seen as a maximum possible deviation for
  the chosen ${\rm N}_{\rm side}$.}
\item
  {\bf  Mask:}
  We have examined the impact of an Euclid type mask on skew-spectrum in a Pseudo-${\cal C}_{\ell}$ based approach introduced in
  \textsection\ref{estimator}.
  The results are presented in Figure - \ref{fig:mask}. The upper solid-lines in each panel correspond to
  to all-sky theoretical predictions of $S_{\ell}$. The upper lines with scatter correspond to
  the estimates from one relaisation of the simulated maps. The left-panel corresponds to the source redshift $z_s=1.0$ and the
  right-panel corresponds to $z_s=2.0$. The lower dashed-curves in each panel correspond to the PCL based theoretical
  predictions $\tilde S_{\ell}$ computed using Eq.(\ref{eq:PCL1}). The corresponding (lower) lines with scatter are estimates from
  one realisation of partial sky with the {\em Euclid}-type mask, shown in Figure-\ref{fig:myhealpymask}, applied. 
\item 
  {\bf  Noise:}
  The impact of noise which we assume to be Gaussian  on estimation of skew-spectrum is shown in Figure - \ref{fig:noise}.
  In both panels the source plane is fixed at $z_s$=1. The solid lines in each panel represent the theoretical skew-spectrum
  for $z_s=1$. The dashed lines represent the pseudo skew-spectrum represented as $\tilde S_l$ with an {\em Euclid}-type mask
  being included. If we compare the scatter with corresponding plots in Figure-\ref{fig:mask} we can
  see how the inclusion of noise increases the scatter though the estimator remains unbiased. 
  skew-spectrum for a Gaussian noise alone is zero so the 
  only effect the noise has on the estimator is to increase its scatter.
  
  The noise was generated at each pixel using a Gaussian deviate with variance $\sigma={\sigma_{\epsilon}/\sqrt{\bar n}}$.
  Where we take $\sigma_{\epsilon}$ represents the variance of the observed ellipticity $\sigma_{\epsilon}=0.3$, and $\bar n$ is the average number
  density of source galaxies per pixel computed using total number of observed galaxies, fraction of sky covered and number of pixel at
  a specific healpix resolution. We have used two different values of $\bar n$.
  The left-panel of Figure-\ref{fig:noise} corresponds to a source density of $n_s=30\; {\rm arcmin}^{-2}$ and the right-panel corresponds to $n_s=10\; {\rm arcmin}^{-2}$.
  The fraction of the sky covered by the mask was taken to be $f_{\rm sky}=0.35$.
\end{enumerate}
%

%
\section{Conclusions and Future Prospects}
\label{sec:conclusion}
%
%
In this paper we have introduced the skew-spectrum statistic as a probe for weak lensing bispectrum.
While we found an excellent agreement of numerical simulations and fitting-function based theoretical predictions
for the auto-correlation we have studied, we also found significant deviation in many other situations and found
that the current analytical uncertainty is not sufficient for high accuracy work.
We have primarily focused on gravity induced
secondary bispectrum in a $\Lambda$CDM cosmology. However, several
extensions of our study are possible.

{\bf Skew-spectrum in beyond $\Lambda$CDM scenarios:} In most modified gravity theories and dark energy models, the bispectrum is currently known only
in the perturbative regime. We have provided analytical expressions for the skew-spectrum in such scenarios.
To go beyond perturbative regime a nonlinear model for the bispectrum is required.
It is expected that a fitting-function based description in such scenarios will eventually be available as more accurate simulations
are performed. Similarly, the modelling of bispectrum based on Effective Field Theories  will also be extended
to modified gravity theories. Once such results are available, they can readily be used to compute the
skew-spectrum in these models.

{\bf Higher-order corrections:} The theoretical expressions of the skew-spectrum are derived using many simplifying assumptions.
We have ignored the corrections due to magnification bias as well as reduced shear which should be included in
more accurate theoretical predictions. In addition the skew-spectrum here is computed using the Limber approximation \citep{Limber}.

{\bf Skew-spectrum from shear maps:} We have computed the skew-spectrum from a convergence map. However,
for many practical purposes a skew-spectrum
estimated directly from shear maps can bypass many of complications of the map making process.

{\bf Intrinsic alignment:} The intrinsic alignment (IA) of galaxies (see \cite{VCS10} and the references therein) are caused by the tidal interaction and is a source of
contamination to gravity induced (extrinsic) weak-lensing. The lensing bispectrum induced
by IA is typically at the level of $10\%$ of the lensing induced bispectrum.
Several methods have been proposed to mitigate or remove such contamination
using joint analysis of power-spectrum and bispectrum. The skew-spectrum  
retains some of the shape information of the original bispectrum.
A joint analysis of power spectrum and skew-spectrum can thus be useful
in separation of these two different contributions. The skew-spectrum
introduced in this study can further be optimised by introducing
weights to judge the level of cross-contamination from the intrinsic
alignment much in the same way as was achieved in case of
point source contamination of CMB studies designed to detect primordial non-Gaussianiaty
from Planck data. 
\

{\bf Primordial non-Gaussianity and active perturbations:} We have considered the gravity induced non-Gaussianity in our study as it is the most dominant
source of non-Gaussianity in weak lensing maps. However, similar results can also obtained
for computing the sub-dominant contributions from primordial non-Gaussianity as well as secondary
sources of non-Gaussianity induced by active sources, e.g., cosmic strings.

{\bf Baryonic Feedback:} We have not included any baryonic feedback in our modelling of the
skew-spectrum but, such corrections can be incorporated in the skew-spectrum for
direct comparison with any realistic data.

  {\bf Covariance and likelihood:} We have not discussed the covariance of the skew-spectrum in this study.
An accurate description of the covariance will be an important ingredient
of cosmological likelihood analysis involving skew-spectrum. A simple form
of covariance can be derived under the assumption of Gaussianity
and thus ignoring all higher-order correlation contributing to the covariance.
Such an estimation will be useful in the noise dominated regime but
will not be sufficient in the highly nonlinear scales characterised by high signal-to-noise
probed by the future surveys such as {\em Euclid}. The methods developed so far
in computing the covariance include the ones based on perturbative analysis,
halo model or simulated mocks \citep{Rizzato}.
These methods can be adapted to compute the skew-spectrum covariance.

\section*{Acknowledgment}
DM is supported by a grant from the Leverhume Trust at MSSL.
It is a pleasure for DM to thank Filippo Vernizzi for useful discussion.
We would like to thank Peter Taylor for providing us his code to generate
the Euclid type mask used in our study as well as for many useful discussions.
We would like to thank Ryuichi Takahashi for careful reading of the draft
and suggestions for improvements. 
\bibliography{SkewSpec.bbl}
\end{document}